\documentclass[11pt, twocolumn]{aastex631}
\usepackage{multirow}

\begin{document}
	
\title{\textbf{\large{
Morphology of Galaxies in JWST Fields: Initial Distribution and Evolution of Galaxy Morphology
}}}

\author[0000-0003-3301-759X]{Jeong Hwan Lee}
\affil{Department of Astronomy and Atmospheric Sciences, Kyungpook National University, Daegu 41566, Republic of Korea}
\affil{The Center for High Energy Physics, Kyungpook National University, Daegu 41566, Republic of Korea}
\author[0000-0001-9521-6397]{Changbom Park}
\affil{Korea Institute for Advanced Study, 85 Hoegi-ro, Dongdaemun-gu, Seoul 02455, Republic of Korea}
\author[0000-0003-3428-7612]{Ho Seong Hwang}
\affiliation{Department of Physics and Astronomy, Seoul National University, 1 Gwanak-ro, Gwanak-gu, Seoul 08826, Republic of Korea}
\affiliation{SNU Astronomy Research Center, Seoul National University, 1 Gwanak-ro, Gwanak-gu, Seoul 08826, Republic of Korea}
\author{Minseong Kwon}
\affiliation{Department of Physics and Astronomy, Seoul National University, 1 Gwanak-ro, Gwanak-gu, Seoul 08826, Republic of Korea}

\correspondingauthor{Ho Seong Hwang}
\email{hhwang@astro.snu.ac.kr}

\begin{abstract}

A recent study from the Horizon Run (HR5) cosmological simulation has predicted that galaxies with ${\rm log}~M_{\ast}/M_{\odot}\lesssim 10$ in the cosmic morning ($10\gtrsim z\gtrsim 4$) dominantly have disk-like morphology in the $\Lambda$CDM universe, which is driven by the tidal torque in the initial matter fluctuations.
For a direct comparison with observation, 
we identify a total of about $19,000$ 
James Webb Space Telescope (JWST) galaxies with ${\rm log}~M_{\ast}/M_{\odot}>9$ at $z=0.6-8.0$ utilizing deep JWST/NIRCam images of publicly released fields, including 
NEP-TDF, NGDEEP, CEERS, COSMOS, UDS, and SMACS J0723$-$7327. 
We estimate their stellar masses and photometric redshifts with the redshift dispersion of 
$\sigma_{\rm NMAD}=0.009$ 
and outlier fraction of only about 6\%.
We classify galaxies into three morphological types, `disks', `spheroids', and `irregulars', applying the same criteria used in the HR5 study.
The morphological distribution of the JWST galaxies shows that disk galaxies account for $60-70\%$ at all redshift ranges.
However, in the high-mass regime (${\rm log}~M_{\ast}/M_{\odot}\gtrsim11$), spheroidal morphology becomes the dominant type.
This implies that mass growth of galaxies is accompanied with morphological transition from disks to spheroids. The fraction of irregulars is about 20\% or less at all mass and redshifts. All the trends in the morphology distribution are consistently found in the six JWST fields. 
These results are in close agreement with the results from the HR5 simulation, particularly confirming the prevalence of disk galaxies at small masses in the cosmic morning and noon.

\end{abstract}


\section{Introduction}

Since the Hubble's tuning fork scheme for galaxy classification was introduced \citep{hub26, san61}, galaxy morphology has provided important insights into the evolution of galaxies across cosmic time.
Previous studies have found that galaxy morphology is closely involved with the environment and various intrinsic properties of galaxies such as luminosity, mass, and star formation activity.
For instance, the morphology-density and morphology-radius relations \citep{dre80, pos05, park07, park09b, fas15} demonstrated that the proportion of early-type galaxies (ellipticals and S0 galaxies) in galaxy clusters increases as clustercentric distance decreases and galaxy number density increases.
These relationships are also dependent on the stellar mass of galaxies, showing significantly high fractions of early-type galaxies ($\gtrsim50\%$) for galaxies more massive than ${\rm log}~M_{\ast}/M_{\odot}>11$ \citep[morphology-mass relation;][]{bam09, vul11, cal12}.
Furthermore, several studies using the Sloan Digital Sky Survey (SDSS) data suggested that the morphology of neighboring galaxies also plays an important role in 
the morphological distribution of galaxies in local universe \citep{park08, park09} and in the Great Observatories Origins Deep Survey (GOODS) field at up to $z\sim1$ \citep{hwang09}.
They showed that the early-type probability of a galaxy increases when the galaxy is closer to another early-type neighbor galaxy, whereas the probability decreases when the neighbor is a late-type galaxy.
In addition, \citet{hwang09} showed that the morphology-density relation becomes much weaker at $z\sim1$ compared to the nearby universe, implying that the morphological distribution of galaxies must depend on redshift.

Beyond $z>1$, high-resolution optical and near-infrared (NIR) images from the Hubble Space Telescope (HST) have been used for investigating the morphology of galaxies.
Most studies utilizing HST data have consistently reported that galaxies at $z\sim2$ exhibit a high prevalence of disturbed or interacting morphologies, which are categorized as peculiar types \citep{abr01, con05, con08, pap05, hue09, hue16, bui13, mor13}.

However, there have been two major problems in previous HST studies on high-redshift galaxy morphology.
First, there are notable discrepancies in the specific trends of morphological distributions.
\citet{bui13} presented that the fraction of disk-like galaxies increases from approximately $15\%$ at $z\sim0$ to $\sim80\%$ at $z>2$, whereas the fraction of spheroid-like galaxies decreases from $\sim85\%$ to $\sim20\%$ over the same redshift range, based on the data from the Palomar Observatory Wide-field InfraRed/DEEP2 (POWIR/DEEP2) and GOODS NICMOS Survey (GNS) surveys.
In contrast, \citet{mor13} found a high fraction of spheroidal galaxies ($40\%$) and a negligible number of disk-like galaxies ($<10\%$) at $z>2$ in the Ultra Deep Survey (UDS) region of the Cosmic Assembly Near-infrared Deep Extragalactic Legacy Survey (CANDELS) field.
These discrepancies could be attributed to variations in the criteria for morphological classification and sample selection.
For instance, \citet{bui13} selected massive galaxies (${\rm log}~M_{\ast}/M_{\odot}>11$) and classified their morphologies mainly based on their S\'ersic indices, while \citet{mor13} only visually classified the morphology of galaxies with ${\rm log}~M_{\ast}/M_{\odot}>10$.
Second, HST optical and NIR data have observational limitations for studying high-redshift galaxy morphology at $z\gtrsim3$.
The longest wavelength coverage of the HST extends to $\sim1.6{\rm\mu m}$ (WFC3-IR/F160W), restricting investigations of galaxy morphology to $z\lesssim2.5$ in the rest-frame optical wavelength range.
Furthermore, the spatial resolution of NIR (WFC3-IR) data is worse than that of HST optical (ACS) data, showing a full width at half maximum (FWHM) twice as broad ($\sim0\farcs2$) as that from the HST/ACS images.
These limitations pose challenges when studying the morphology of high-redshift galaxies.

Thanks to the unprecedented sensitivity and resolution of the James Webb Space Telescope (JWST), there have been dramatic advances in observational studies of high-redshift galaxy morphology.
Recent JWST studies have utilized various methods for classifying galaxy morphology, including visual inspections \citep{fer22, fer23, jac23, kar23}, supervised Convolutional Neural Networks \citep{hue23}, the Morpheus deep-learning framework \citep{rob23}, and unsupervised machine learning \citep{toh23, veg24}.
\citet{fer22,fer23} conducted visual classifications of galaxy morphology combined with nonparametric measurements of concentration, asymmetry, and smoothness \citep[CAS;][]{con03}, using JWST/NIRCam images of the SMACS 0723 and the Cosmic Evolution Early Release Science Survey (CEERS) fields.
They concluded that disk galaxies are predominant at $z=3-6$, constituting $>40\%$ of galaxies with ${\rm log}~M_{\ast}/M_{\odot}>9$.
With a sample of 850 galaxies from the CEERS field, \citet{kar23} also found a high fraction of disk galaxies, comprising $60\%$ at $z\sim3$ and $30\%$ at $z>6$, while the fractions of spheroid and irregular galaxies are relatively constant at all redshifts.
\citet{hue23} and \citet{toh23} investigated the morphology of a relatively larger sample of galaxies at $z>2$ by employing artificial intelligence techniques.
They demonstrated that the fractions of clumpy and irregular galaxies increase with increasing redshift, while the fractions of spheroid-like galaxies decrease. 
\citet{veg24} explored the morphological distribution of high-redshift galaxies at $z\sim3-6$ using a contrastive learning framework.
They found that the fraction of disk galaxies might be overestimated, attributed to the misclassification of compact prolate-shaped galaxies as disks.
Overall, recent JWST studies are in agreement that the Hubble Sequence, encompassing diverse morphologies from disk galaxies to spheroid galaxies, was already established in the early universe at $z\sim8-9$.
This agreement results from the superior observational performance of the JWST, which have improved the visibility of regular morphological features such as disks, bulges, and spiral arms compared to HST \citep{jac23}.

\begin{figure}
\centering
\includegraphics[width=0.5\textwidth]{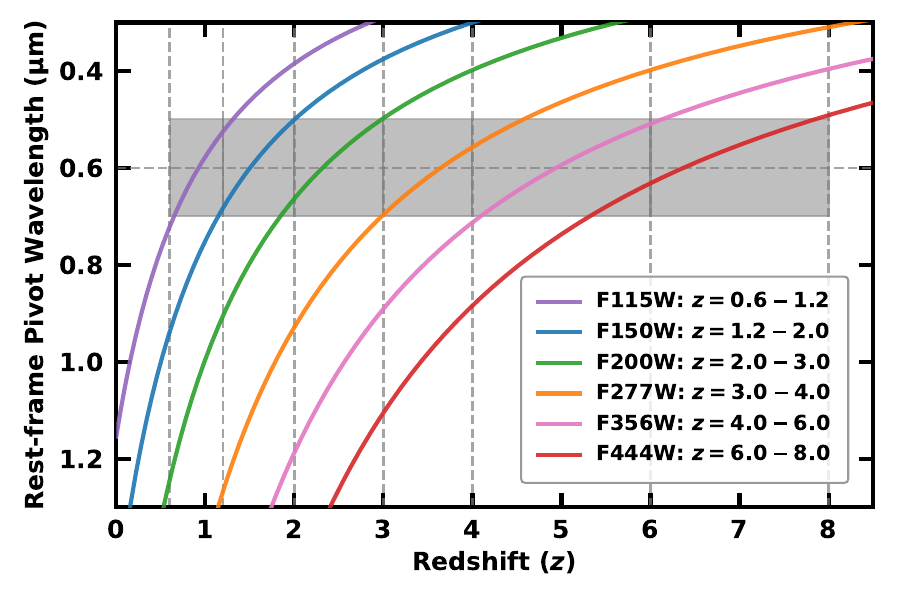}
\caption{
Rest-frame pivot wavelengths of JWST/NIRCam filters (F115W, F150W, F200W, F277W, F356W, and F444W) as a function of redshift.
The shaded gray region represents the rest-frame optical wavelength range spanning $0.5-0.7~{\rm \mu m}$.
\label{fig:wav}}
\end{figure}

In this work, we examine the rest-frame optical morphologies of a large sample of $\sim19,000$ galaxies at $z=0.6-8.0$ using data from six public fields of JWST. We have improved the accuracy of photometric redshifts significantly, which can potentially impact the study of redshift evolution of galaxy morphology.
{\color{blue} \bf Figure \ref{fig:wav}} displays the rest-frame pivot wavelengths of JWST/NIRCam filters (F115W, F150W, F200W, F277W, F356W, and F444W) as a function of redshift.
To probe the galaxy morphology, we choose the NIRCam filter to observe the galaxies in rest-frame optical wavelengths ($0.5-0.7~{\rm \mu m}$) at the corresponding redshifts.
We then compare these galaxy morphologies with the recent findings from Horizon Run 5 (HR5) cosmological simulation \citep{lee21}.
The HR5 results from \citet{park22} suggested that approximately two-thirds of the galaxies with $M_{\ast}\leq 10^{10}~M_{\odot}$ in the cosmic morning ($10\gtrsim z\gtrsim4$) exhibit disk-like morphologies, implying the dominance of disks in the early universe. The remaining galaxies are roughly equally divided into irregulars and spheroids, with a slight prevalence of irregulars.
Our objective is to verify if these HR5 results are supported by observational evidence, using high-redshift galaxies detected in JWST/NIRCam images when the morphology classification criteria as close as those used in HR5 are applied.
To achieve this scientific goal, we selected galaxies with stellar mass of $M_{\ast}>10^{9}~M_{\odot}$ across a wide range of redshifts ($z=0.6-8.0$).
In line with the HR5 study, we applied the same morphology classification scheme as depicted in Figure 2 of \citet{park22}.

\begin{deluxetable*}{ccccc}
	\tabletypesize{\footnotesize}
	\setlength{\tabcolsep}{0.11in}
	\tablecaption{HST and JWST Images and Filters Used in This Study}
	\tablehead{\colhead{Field} & \colhead{Telescope} & \colhead{Instrument} & \colhead{Filters} & 
    \colhead{Coverage Area\tablenotemark{\rm \footnotesize a}} \\
     & & & & (${\rm arcmin^{2}}$)}
	\startdata
	NEP-TDF & HST & ACS & F435W, F606W & 22.0 \\
     & JWST & NIRCam (SWC) & F090W, F115W, F150W, F200W & 10.5 \\
     & & NIRCam (LWC) & F277W, F356W, F410M, F444W & 10.0 \\ \hline
	NGDEEP & HST & ACS & F435W, F606W, F775W, F814W, F850LP & 37.3 \\
     & & WFC3-IR & F105W, F125W, F160W & 30.7 \\
     & JWST & NIRCam (SWC) & F115W, F150W, F200W & 11.2 \\
     & & NIRCam (LWC) & F277W, F356W, F444W & 9.4 \\ \hline  
	CEERS & HST & ACS & F435W, F606W, F814W & 201.3 \\
     & & WFC3-IR & F125W, F140W, F160W & 167.4 \\
     & JWST & NIRCam (SWC) & F115W, F150W, F200W & 89.9 \\
     & & NIRCam (LWC) & F277W, F356W, F410M, F444W & 91.7 \\ \hline
	COSMOS & HST & ACS & F435W, F475W, F606W, F814W & 245.6 \\
     & & WFC3-IR & F125W, F140W, F160W & 241.3 \\
     & JWST & NIRCam (SWC) & F090W, F115W, F150W, F200W & 86.8 \\
     & & NIRCam (LWC) & F277W, F356W, F410M, F444W & 88.9 \\ \hline
	UDS & HST & ACS & F435W, F606W, F814W & 202.8 \\
     & & WFC3-IR & F125W, F140W, F160W & 186.6 \\
     & JWST & NIRCam (SWC) & F090W, F115W, F150W, F200W & 147.4 \\
     & & NIRCam (LWC) & F277W, F356W, F410M, F444W & 150.8 \\ \hline
	SMACS0723 & HST & ACS & F606W, F814W & 12.3 \\
     & & WFC3-IR & F105W, F125W, F140W, F160W & 5.5 \\
     & JWST & NIRCam (SWC) & F090W, F150W, F200W & 11.4 \\
     & & NIRCam (LWC) & F277W, F356W, F444W & 11.0 \\
	\enddata
	\label{tab:imgs}
	\tablenotetext{}{\textbf{Note.}}
	\tablenotetext{\rm a}{Areas are measured based on the image observed by the instrument's filter with the longest wavelength.}
\end{deluxetable*}

This paper is structured as follows.
In Section \ref{sec:img}, we explain the HST and JWST images used in our analysis.
Sections \ref{sec:phot} and \ref{sec:sed} describe how we conducted multiwavelength photometry and estimated photometric redshifts and stellar masses of galaxies through spectral energy distribution (SED) fitting.
Section \ref{sec:morp} describes the application of the galaxy classification scheme, with measurements of S\'ersic indices and asymmetry.
In Section \ref{sec:result}, we illustrate our major findings about the morphological fractions as a function of redshift and stellar mass.
In Section \ref{sec:discuss}, we discuss the implications of our results, including a comparison with the HR5 study.
Finally, Section \ref{sec:summary} summarizes our key results.
Throughout this paper, we adopt the cosmological parameters in the Planck 2015 results \citep{plc16}, which are also employed in the HR5 simulation \citep{lee21}: $H_{0}=68.4~{\rm km~s^{-1}~Mpc^{-1}}$, $\Omega_{M}=0.3$, $\Omega_{\Lambda}=0.7$, $\Omega_{\rm b}=0.047$, and $\sigma_{8}=0.816$.

\section{Data and Methods}
\label{sec:dat+met}

\subsection{HST and JWST/NIRCam Images}
\label{sec:img}

In this study, we made use of publicly available images of HST and JWST/NIRCam from the Grizli Image Release v6.0\footnote[1]{\url{https://s3.amazonaws.com/grizli-v2/JwstMosaics/v6/index.html}\\
\url{https://grizli.readthedocs.io/en/stable/grizli/image-release-v6.html}}, 
which was reduced and processed by the \textsc{Grizli} pipeline \citep{bra23a}.
This pipeline has been known to be effective to alleviate instrumental artifacts in JWST countrate products (\texttt{*\_rate.fits}), including vertical and horizontal stripes from electronic read noise \citep[commonly known as `$1/f$ noise';][]{sch20}, circular patterns by significant cosmic ray events (`snowballs'), and stray light features like `wisps' and `claws' \citep{rig23}.
After correcting these artifacts, the pipeline aligned the HST and JWST/NIRCam images using stars from GAIA DR3 catalog \citep{gaia21}, and combined those images.
Detailed information about the image reduction process was provided in previous JWST studies \citep{bez22, brad23, val23}.
The final mosaic images are sampled at pixel scales of $0\farcs04~{\rm pixel^{-1}}$ for HST and JWST/NIRCam long wavelength channels (F277W, F356W, and F444W) and $0\farcs02~{\rm pixel^{-1}}$ for JWST/NIRCam short wavelength channels (F090W, F115W, F150W, and F200W).
For our analysis, we resampled the NIRCam short-wavelength images to $0\farcs04~{\rm pixel^{-1}}$ using the Python-based tool \texttt{astropy reproject}\footnote[2]{\url{https://github.com/astropy/reproject}}.

We obtained the processed HST and JWST images of six public fields: the North Ecliptic Pole Time-Domain Fields \citep[NEP-TDF;][]{win22, win23}, the Next Generation Deep Extragalactic Exploratory Public (NGDEEP) Survey \citep{fin21, bag23}, the CEERS \citep{fin17, fin23}, the Cosmic
Evolution Survey \citep[COSMOS;][]{dun21, kar21, cas23}, the UKIRT Infrared Deep Sky Survey Ultra-deep Survey field \citep[UDS;][]{dun21}, and the SMACS J0723.3-7327 cluster \citep[SMACS0723;][]{pon22}.
{\color{blue} \bf Table \ref{tab:imgs}} summarizes the HST and JWST images used in our spectral energy distribution (SED) analysis, excluding images from filters with either excessively narrow coverage areas (e.g., HST F105W in most fields) or shallow depths (e.g., HST F275W in CEERS and HST F435W in SMACS0723).

\begin{figure*}
\centering
\includegraphics[width=0.9\textwidth]{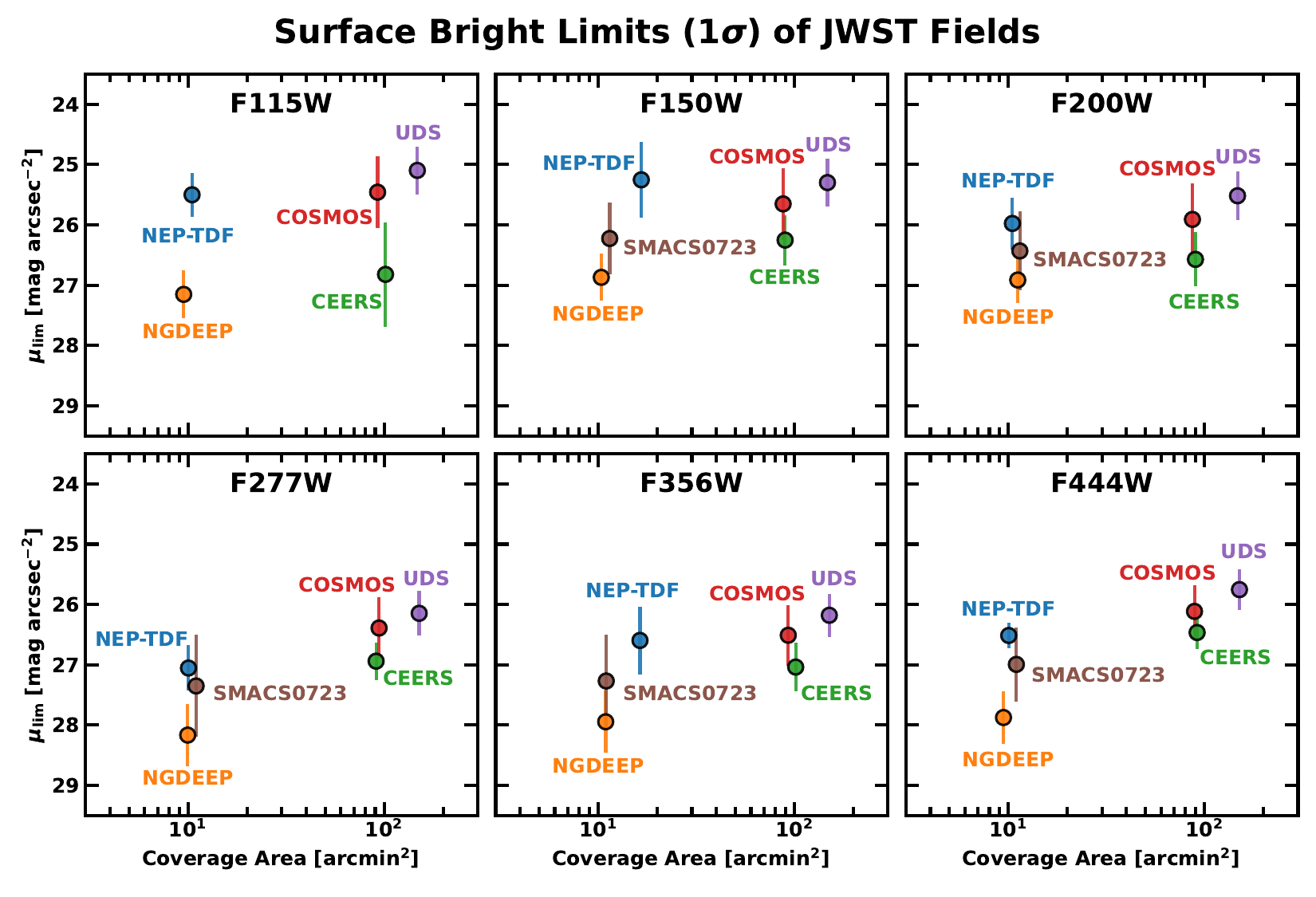}
\caption{
Distributions of surface brightness limits and coverage areas for JWST NIRCam filters (F115W, F150W, F200W, F277W, F356W, and F444W). Surface brightness limits are measured within apertures with a radius of $0\farcs5$.
Note that the SMACS0723 field has no JWST images from F115W.
\label{fig:sblim}}
\end{figure*}

{\color{blue} \bf Figure \ref{fig:sblim}} illustrates the surface brightness limit and coverage area of each JWST field in various NIRCam filters.
We measured the $1\sigma$ surface brightness levels above the background within circular apertures with a radius of $0\farcs5$.
For these measurements, we randomly selected 100 circular apertures per 
field of view of $2\farcm2\times2\farcm2$, corresponding to a single JWST/NIRCam module.
Overall, the JWST NIRCam images exhibit surface brightness limits deeper than $\sim25~{\rm mag~arcsec^{-2}}$ for the short-wavelength filters and $\sim26~{\rm mag~arcsec^{-2}}$ for the long-wavelength filters, which even enables the detection of low surface brightness galaxies \citep{ike23}.
The NGDEEP ($\mu_{\rm lim}({\rm F444W})\sim28.0~{\rm mag~arcsec^{-2}}$) and UDS (coverage of $\sim 150~{\rm arcmin^{2}}$) are the deepest and widest fields in this study, respectively.

\subsection{Multiwavelength Photometry}
\label{sec:phot}

\begin{figure*}
\centering
\includegraphics[width=0.9\textwidth]{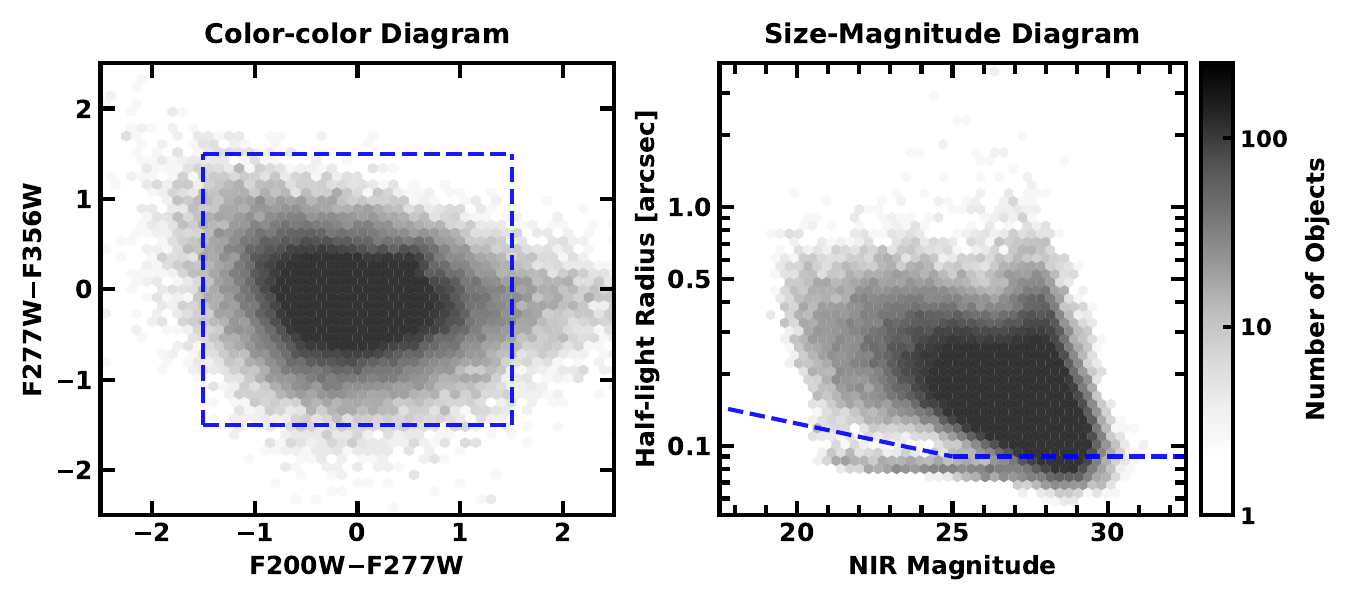}
\caption{
Left panel: Color-color diagram of the detected sources with $\texttt{FLAGS}\leq4$ across all JWST fields.
Measurements of ${\rm F200W}-{\rm F277W}$ and ${\rm F277W}-{\rm F356W}$ were conducted within the Kron radii of the objects.
The blue dashed box denotes the selection criteria of $-1.5<{\rm F200W}-{\rm F277W}<1.5$ and $-1.5<{\rm F277W}-{\rm F356W}<1.5$ to discard artifacts.
Right panel: Size-magnitude diagram of the sources, with half-light radii (\texttt{FLUX\_RADIUS}) and magnitudes measured in the reference image (${ \rm F277W}+{\rm F356W}+{\rm F444W}$).
The blue dashed line represents the boundary between extended sources and point sources.
Point source candidates located below this boundary were excluded from our sample for the analysis.
\label{fig:phots}}
\end{figure*}

For target selection and SED analysis for high-redshift galaxies, we performed source detection and aperture photometry on HST and JWST images using \textsc{Source Extractor} version 2.25.0\footnote[3]{\url{https://github.com/astromatic/sextractor}} \citep[\textsc{SExtractor};][]{ber96}.
We employed the combined images of the NIRCam long-wavelength filters (F277W+F356W+F444W) as the reference images for performing forced photometry on both HST and JWST images.
We utilized a two-mode detection process with `cold' and `hot' modes as applied in 
\citet{gal13}.
The cold-mode configurations were adjusted to prevent overdeblending of large and bright sources, whereas the hot-mode configurations were set to detect faint objects close to bright sources.
{\color{blue} \bf Table \ref{tab:sep}} lists the input configuration parameters for the cold-mode and hot-mode detections of \textsc{SExtractor}.
In the case of SMACS0723 cluster, we adjusted the configuration parameters slightly to obtain sharper background estimation due to intense intracluster light.
Subsequently, we combined the source catalogs from the two-mode photometry, by retaining all sources from the cold mode and supplementing them with hot-mode sources located outside the Kron radii of cold-mode sources \citep{bar12, kar23}.
The merged source catalogs include various objects other than galaxies such as cosmic rays, artifacts, or foreground stars.
Therefore, 
we initially selected galaxy candidates using specific criteria based on colors and sizes as shown in {\color{blue} \textbf{Figure \ref{fig:phots}}}.
We applied realistic color ranges ($-1.5<{\rm F200W-F277W}<1.5$ and $-1.5<{\rm F277W-F356W}<1.5$) and $\texttt{FLAGS}\leq4$ to remove artifacts.
These color ranges were chosen to reflect the typical JWST/NIRCam colors of stellar populations within galaxies, and the \texttt{FLAGS} effectively filtered out saturated sources or those close to image edges.
Furthermore, to minimize contamination from cosmic rays and foreground stars, 
we rejected point sources with compact half-light radii ($\texttt{FLUX\_RADIUS}\leq0\farcs09$ or $\texttt{FLUX\_RADIUS}\leq-0.008\times(\texttt{MAG\_AUTO}-25)+0\farcs09$) in the reference images.
The size criterion was visually determined on the size-magnitude diagram, which follows the horizontal sequence of point sources, with a more stringent cut applied for bright objects due to the presence of saturated stars.
The numbers of all detected sources and initially selected galaxy candidates are listed in the first and second columns of {\color{blue} \textbf{Table \ref{tab:samp}}}.

\begin{deluxetable*}{ccc}
	\tabletypesize{\footnotesize}
	\setlength{\tabcolsep}{0.25in}
	\tablecaption{\textsc{SExtractor} Input Parameters Used for Multiwavelength Photometry}
	\tablehead{\colhead{Parameter} & \colhead{Cold-mode Configuration} & \colhead{Hot-mode Configuration}}
	\startdata
    \texttt{DETECT\_MINAREA} & 20 & 20 \\
    \texttt{DETECT\_THRESH} & 1.0 & 1.0 \\
    \texttt{ANALYSIS\_THRESH} & 1.0 & 1.0 \\
    \texttt{FILTER\_NAME} & tophat\_5.0\_5x5.conv & gauss\_2.5\_5x5.conv \\
    \texttt{DEBLEND\_NTHRESH} & 32 & 16 \\
    \texttt{DEBLEND\_MINCONT} & 0.01 & 0.0001 \\
    \texttt{SATUR\_LEVEL} & 100.0 & 100.0 \\
    \texttt{BACK\_SIZE\tablenotemark{\rm \footnotesize a}} & 64 (32) & 16 \\
    \texttt{BACK\_FILTERSIZE\tablenotemark{\rm \footnotesize a}} & 5 (3) & 3 \\
    \texttt{BACK\_PHOTOTHICK\tablenotemark{\rm \footnotesize a}} & 48 (24) & 24 (12) \\
	\enddata
	\label{tab:sep}
	\tablenotetext{}{\textbf{Note.}}
	\tablenotetext{\rm a}{Parameter values within parentheses are applied for the SMACS0723 field.}
\end{deluxetable*}

\subsection{Photometric Redshifts and Stellar Masses}
\label{sec:sed}

\subsubsection{SED Fitting Based on eazy-py}

Prior to conducting SED fitting for galaxy candidates, we carried out an aperture correction procedure to determine the total flux of each source, following the method described in \citet{val23}.
We measured the enclosed flux within a circular aperture with a radius of $0\farcs5$ and accounted for additional flux beyond the $0\farcs5$ aperture within the Kron radius \citep{kro80}.
These aperture-corrected flux values were computed from the reference images and were subsequently applied to all HST and JWST filters.
We employed these aperture-corrected total fluxes for the SED analysis described in Section \ref{sec:sed}.
To correct galactic extinctions in all HST and JWST bands, we made use of \textsc{dustmaps} \citep{gre18}, with adopting a standard $R_{V}=3.1$ and dust reddening maps provided in \citet{sch98}.

\begin{figure*}
\centering
\includegraphics[width=0.75\textwidth]{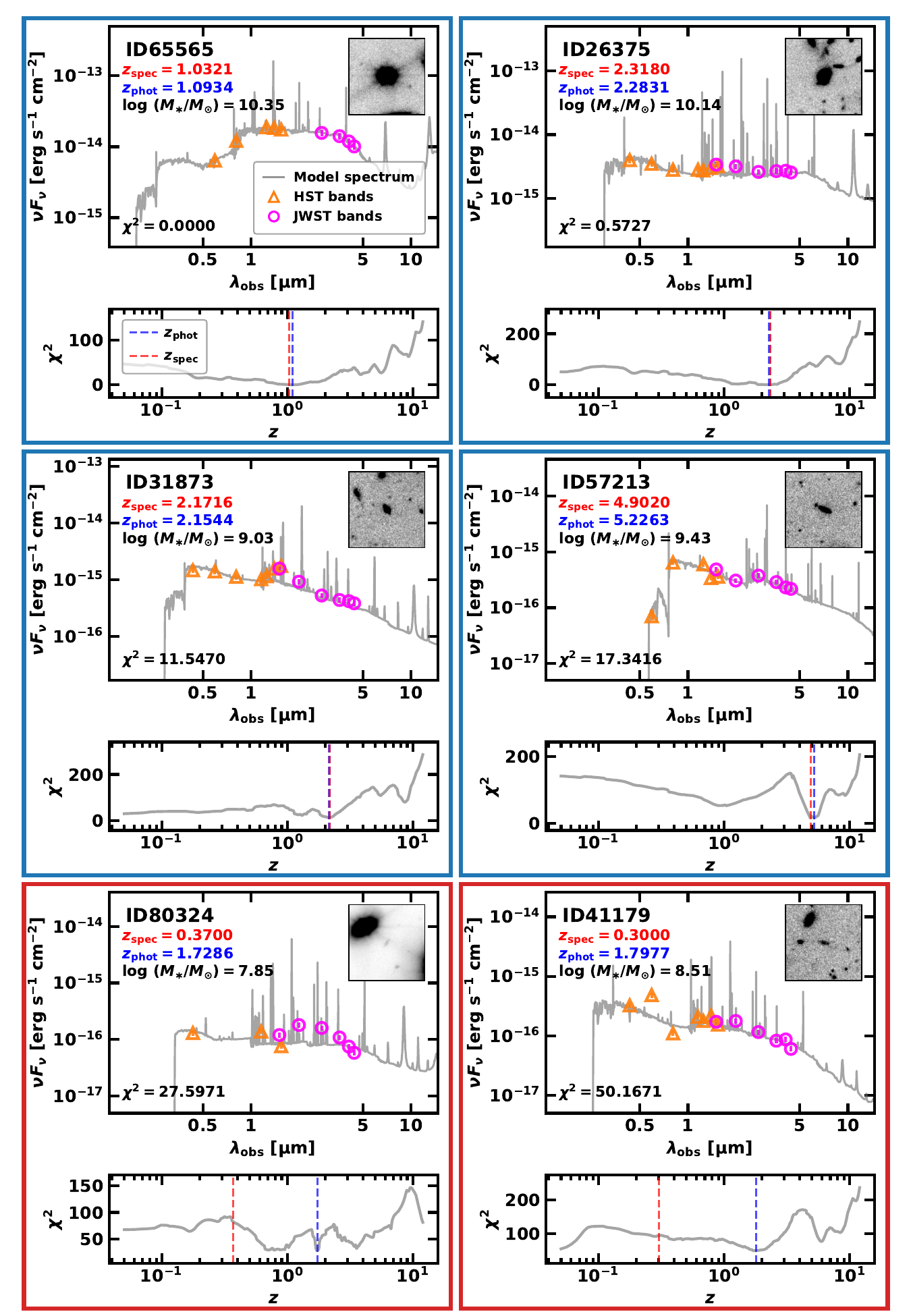}
\caption{
Examples of the SED fitting results for six galaxies in the CEERS field, ordered by their chi-square values of the best-fit \textsc{eazy-py} models.
In this figure, there are three panels for each galaxy.
The upper panel in each set displays the SED of the best-fit model (gray solid line) and photometric data from HST (orange triangles) and JWST (magenta circles) observations.
The lower panel shows the chi-square values as a function of redshift.
The spectroscopic redshift and the photometric redshift are depicted as red and blue dashed lines, respectively.
A small panel on the right side displays the observed galaxy in the reference image (F277W+F356W+F444W).
Based on a criterion of $\chi^{2}=20$, four galaxies in blue boxes are considered to show secure SED fitting results, while two galaxies in red boxes are thought to show unreliable results.
\label{fig:sedfit}}
\end{figure*}

Photometric redshifts and stellar masses of the galaxy candidates
are estimated through the SED fitting process using the \textsc{eazy-py}\footnote[4]{\url{https://github.com/gbrammer/eazy-py}} package \citep{bra08, bra23b}.
Multiwavelength fluxes measured in HST and JWST bands (see {\color{blue} \textbf{Table \ref{tab:imgs}}}) were used for fitting the \textsc{eazy-py} model.
We set the flux uncertainties to a minimum error floor of $10\%$ to allow some flexibility in the SED fitting.
We employed 14 spectral templates from the \texttt{sfhz} model\footnote[5]{\url{https://github.com/gbrammer/eazy-photoz/tree/master/templates/sfhz}}, including 13 templates in the \texttt{corr\_sfhz\_13} subsets and an additional best-fit template from a strong emission-line galaxy observed in the SMACS0723 field \citep{car23}.
Photometric redshifts were determined by identifying the redshift value corresponding to the minimum chi-sqaure value within the range from $z=0.05$ to $z=12.0$, with an interval of ${\rm log}~(1+z)=0.005$.
An iterative correction process for flux zeropoints was performed over 10 steps to optimize redshift values.
No priors were applied on apparent magnitude and ultraviolet (UV) slope during this procedure.

After completing the SED fitting process, we evaluated the reliability of the best-fit SED models by utilizing the chi-square values ($\chi^{2}_{\rm SED}$) from \textsc{eazy-py} and the number of filters ($N_{\rm filt}$) used for the SED analysis.
Through visual inspection of the SED fitting results, we employed a criterion of $\chi^{2}_{\rm SED}<20$ and $N_{\rm filt}\geq 6$ to select well-fitted galaxy candidates.
It is worth noting that our chi-square criterion is more rigorous compared to that applied in \citet{val23}, which used $\chi^{2}_{\rm SED}/N_{\rm filt}\leq8$ and $N_{\rm filt}\geq6$.
{\color{blue} \textbf{Figure \ref{fig:sedfit}}} illustrates examples of SED fitting results for six galaxies located in the CEERS field.
Among these, four galaxies with $\chi^{2}_{\rm SED}<20$ exhibit well-matched SEDs with their photometric data.
In contrast, the best-fit SEDs for two galaxies with $\chi^{2}_{\rm SED}\geq20$ do not align with the photometric data.
These poor results might be associated with contamination from a neighboring bright galaxy (ID 80324) and the faintness of the galaxy itself (ID 41179), as seen in visual representations of the galaxies in the figure.
For further analysis, we decided to exclude samples with $\chi^{2}_{\rm SED}\geq20$ or $N_{\rm filt}<6$ from the \textsc{eazy-py} SED fitting.

\begin{figure*}
\centering
\includegraphics[width=0.9\textwidth]{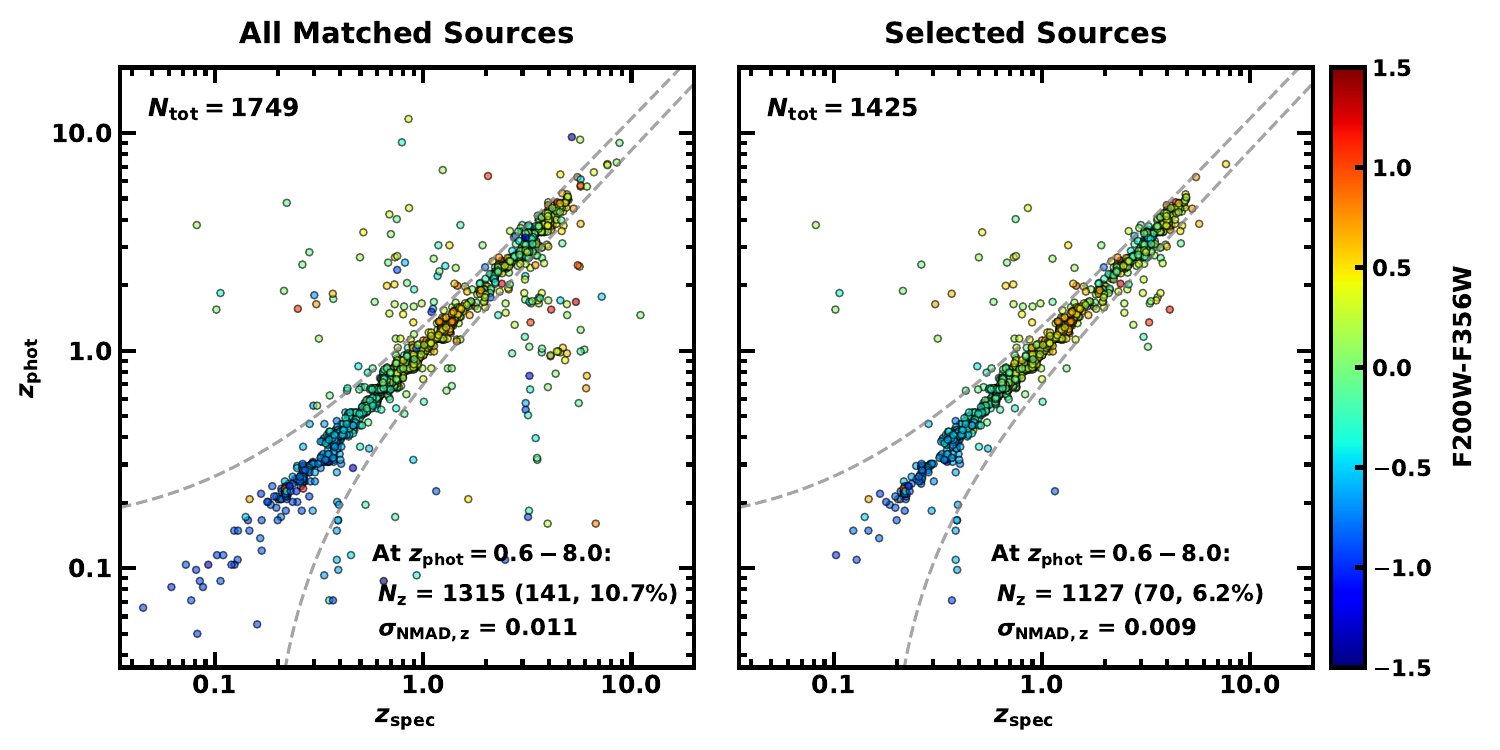}
\caption{
Comparison of photometric redshifts ($z_{\rm phot}$) derived from our \textsc{eazy-py} model in this study with spectroscopic redshifts ($z_{\rm spec}$) obtained from the NED and the literature.
The left panel displays the comparison results for all sources, and the right panel displays the results for sources selected with ${\rm log}~M_{\ast}/M_{\odot}>9$, $\chi^{2}_{\rm SED}<20$, and $N_{\rm filt}\geq6$.
Gray dashed lines indicate the criterion of $|z_{\rm spec}-z_{\rm phot}|/(1+z_{\rm spec})=0.15$, which is used to identify outliers.
This figure provides the total numbers of sources, 
the numbers of outliers, 
the outlier fractions, 
and the $\sigma_{\rm NMAD}$ values as text labels.
\label{fig:zpzs}}
\end{figure*}

\subsubsection{Comparisons of Photometric and Spectroscopic Redshifts}

In this section, we compare photometric redshifts with spectroscopic redshifts to evaluate the reliability of our photometric redshift estimations.
The spectroscopic redshift data for galaxies within JWST fields were obtained from 
\citet{https://doi.org/10.26132/ned1} 
and the available literature.
For all JWST fields, we initially obtained spectroscopic redshifts through a cone search in NED\footnote[6]{\url{https://ned.ipac.caltech.edu/conesearch}}.
Then, the spectroscopic redshifts of galaxies in the CEERS and SMACS0723 fields were additionally sourced from \citet{nak23} and \citet{noi23}, which provided the redshift catalogs utilizing JWST/NIRSpec, JWST/NIRISS grism, and VLT/MUSE data.
These spectroscopic sources were spatially matched with our photometric sources with ${\rm F200W}<27~{\rm mag}$, allowing for a tolerance of $1\farcs5$.
This tolerance was determined to consider potential offsets in the world coordinate system (WCS) between JWST images and previous data.
In the NEP-TDF field, there was only one object with measured spectroscopic redshift (WISEA J172302.06+654802.9; $z_{\rm spec}=0.179$), which is statistically insufficient for redshift comparisons.

{\color{blue} \textbf{Figure \ref{fig:zpzs}}} displays the comparisons of photometric and spectroscopic redshifts from five JWST fields except for NEP-TDF.
We plot the results from all matched sources in the left panel and those from the sources with stellar mass higher than ${\rm log}~M_{\ast}/M_{\odot}>9$, low chi-square values ($\chi^{2}_{\rm SED}<20$), and a minimum six available filters ($N_{\rm filt}\geq6$) in the right panel.
Outliers in this comparison were identified using the criterion of $|z_{\rm spec}-z_{\rm phot}|/(1+z_{\rm spec})>0.15$ \citep{wea23}, which are located outside the gray dashed lines in the figure.
To assess the accuracy of photometric redshifts, we also computed the normalized median absolute deviation ($\sigma_{\rm NMAD}$) of the redshift differences ($\Delta z=z_{\rm phot}-z_{\rm spec}$) as defined in Equation 7 of \citet{bra08}.
For all matched sources, the photometric redshifts of galaxy candidates show a good agreement with their spectroscopic redshifts in the redshift range from $z\sim0.1$ to $z\sim10$, with an outlier fraction about $11\%$ and $\sigma_{\rm NMAD}=0.011$.
When the stricter criterion of ${\rm log}~M_{\ast}/M_{\odot}>9$, $\chi^{2}_{\rm SED}<20$, and $N_{\rm filt}\geq6$ is applied, the results become more reliable, yielding a lower outlier fraction of $\sim6\%$.
The performance of photometric redshift estimation slightly varies with fields, as detailed in {\color{blue} \textbf{Table \ref{tab:zpzs}}}.
The outlier fractions and $\sigma_{\rm NMAD}$ values from our analysis are lower than those in previous JWST studies, which reported an outlier fraction of $10.8\%$ and $\sigma_{\rm NMAD}=0.03$ in the UNCOVER survey field \citep{wea23}, and $\sigma_{\rm NMAD}=0.018$ in the CEERS field \citep{val23}.
This implies that our sample is more rigorously selected based on the SED fitting results than in previous studies.

\begin{deluxetable*}{ccccccccc}
	\tabletypesize{\footnotesize}
	\setlength{\tabcolsep}{0.11in}
	\tablecaption{Comparison Results of $z_{\rm phot}$ versus $z_{\rm spec}$ in Each JWST Field\tablenotemark{\rm \footnotesize a, b}}
	\tablehead{\colhead{Field} & \multicolumn{2}{c}{Number of Matched Sources} & \multicolumn{2}{c}{Number of Outliers} & \multicolumn{2}{c}{Outlier Fraction (\%)} & \multicolumn{2}{c}{$\sigma_{\rm NMAD}$} \\
    & {All $z$} & {Selected $z$} & {All $z$} & {Selected $z$} & {All $z$} & {Selected $z$}}   
	\startdata
    NGDEEP & 62 (48) & 42 (37) & 7 (3) & 4 (3) & 11.3 (6.3) & 9.5 (8.1) & 0.007 \\
    CEERS & 390 (295) & 239 (207) & 43 (15) & 31 (13) & 11.0 (5.1) & 13.0 (6.3) & 0.009 \\
    COSMOS & 627 (531) & 460 (410) & 48 (23) & 42 (22) & 7.7 (4.3) & 9.1 (5.4) & 0.009 \\
    UDS & 485 (392) & 463 (384) & 63 (29) & 54 (29) & 13.0 (7.4) & 11.7 (7.6) & 0.012 \\
    SMACS0723 & 185 (159) & 72 (63) & 16 (8) & 7 (3) & 8.6 (5.0) & 9.7 (4.8) & 0.011 \\ \hline
    Total & 1,749 (1,425) & 1,276 (1,101) & 177 (78) & 138 (70) & 10.1 (5.5) & 10.8 (6.4) & 0.009
	\enddata
	\label{tab:zpzs}
	\tablenotetext{}{\textbf{Notes.}}
	\tablenotetext{\rm a}{Values in parentheses correspond to results with ${\rm log}~M_{\ast}/M_{\odot}>9$, $\chi^{2}<20$, and $N_{\rm filt}\geq6$.}
	\tablenotetext{\rm b}{The selected range of redshift is $z=1.2-8.0$ for SMACS0723 and $z=0.6-8.0$ for the other fields.}
\end{deluxetable*}

However, it is notable that there are several sources at $z_{\rm spec}\lesssim1$ showing overestimated photometric redshifts with $z_{\rm phot}\gtrsim1$.
This discrepancy can be attributed to contamination from adjacent light sources, such as blending with other galaxy light, bright background features, or bleeding light from saturated stars.
These overestimated cases could be low-redshift contaminants in our analysis.
To minimize these contaminants, we manually excluded a few sources close to saturated stars and bright galaxies for all JWST fields through visual inspections.

\begin{figure*}
\centering
\includegraphics[width=0.9\textwidth]{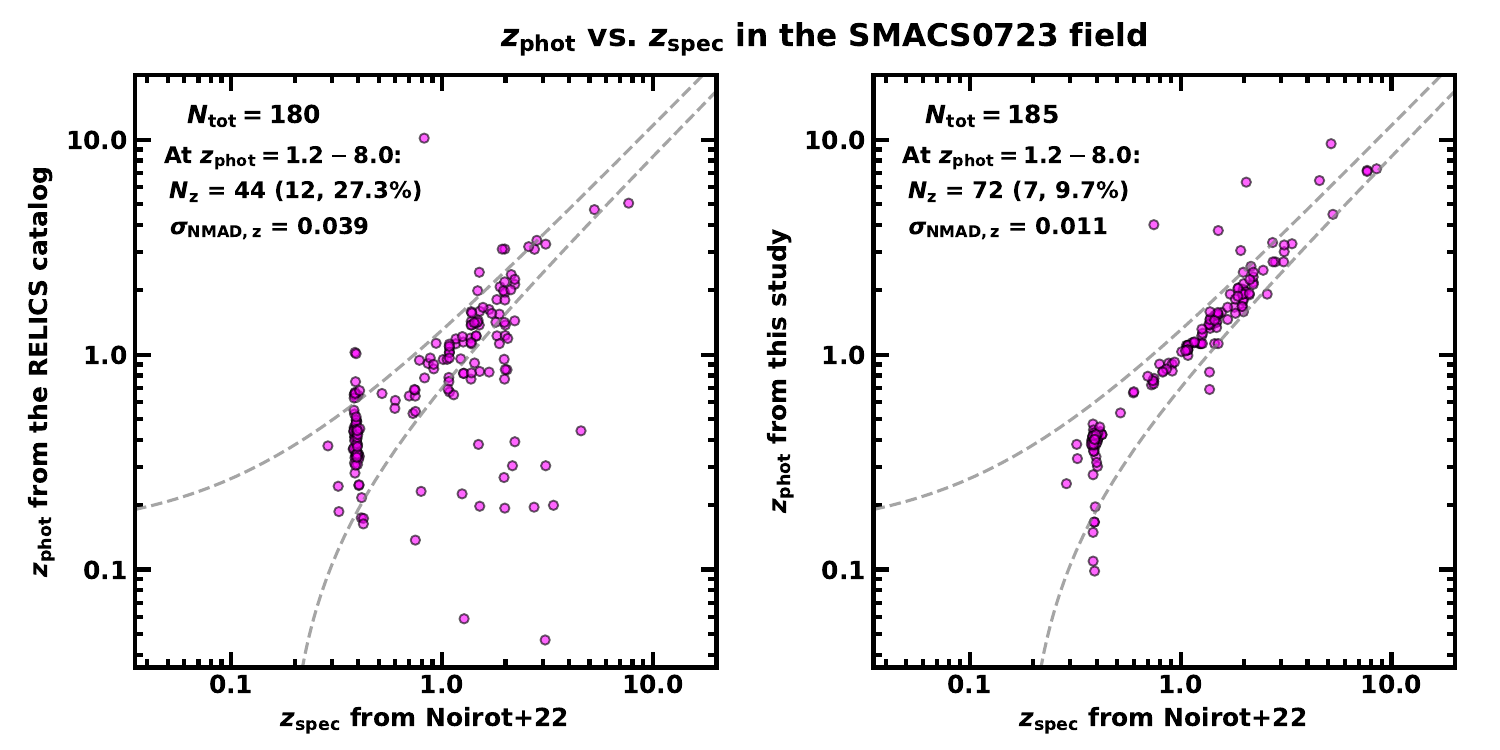}
\caption{
Comparisons of $z_{\rm phot}$ derived from the EAZY-PY model (left panel; this study) and the RELICS catalog \citep[right panel;][]{coe19} 
adopted by \citet{fer22}
with $z_{\rm spec}$ \citep{nak23, noi23}, for the sources detected in the SMACS0723 field.
Gray and purple lines and texts are the same as {\color{blue} {\bf Figure \ref{fig:zpzs}}}.
\label{fig:zpzs_relics}}
\end{figure*}

In addition, we also compared the accuracy of photometric redshifts of galaxies in the SMACS0723 field with those listed in the RELICS catalog \citep{coe19}.
The RELICS study derived photometric redshifts using the Bayesian
photoz code \citep[BPz;][]{ben00}, with HST photometric data from optical to NIR wavelengths.
These photometric redshifts were adopted by \citet{fer22} for investigating the morphologies of high-redshift galaxies.
{\color{blue} \textbf{Figure \ref{fig:zpzs_relics}}} represents the comparisons of $z_{\rm phot}$ vs. $z_{\rm spec}$ both for this study (left panel) and the RELICS catalog (right panel).
The RELICS sources have 180 matched objects with the spectroscopic sample from \citet{nak23} and \citet{noi23}, which are similar to this study.
In comparison to the RELICS catalog, our photometric redshift measurements demonstrate a superior match with spectroscopic redshifts, showing a lower outlier fraction and $\sigma_{\rm NMAD}$.
On the other hand, the RELICS catalog tends to underestimate redshifts of some galaxies at $z_{\rm spec}>1$, which might be due to observational limitations in HST NIR data capturing the redshifted Balmer breaks of these galaxies.
These suggest that some 
high-redshift galaxy candidates in the SMACS0723 field might have been missed in \citet{fer22} because of underestimation of photometric redshifts in the RELICS data.
Furthermore, our analysis demonstrates that the inclusion of JWST NIRCam photometric data improves the accuracy of photometric redshifts when compared to the results obtained using HST-only data.

\begin{figure}
\centering
\includegraphics[width=0.5\textwidth]{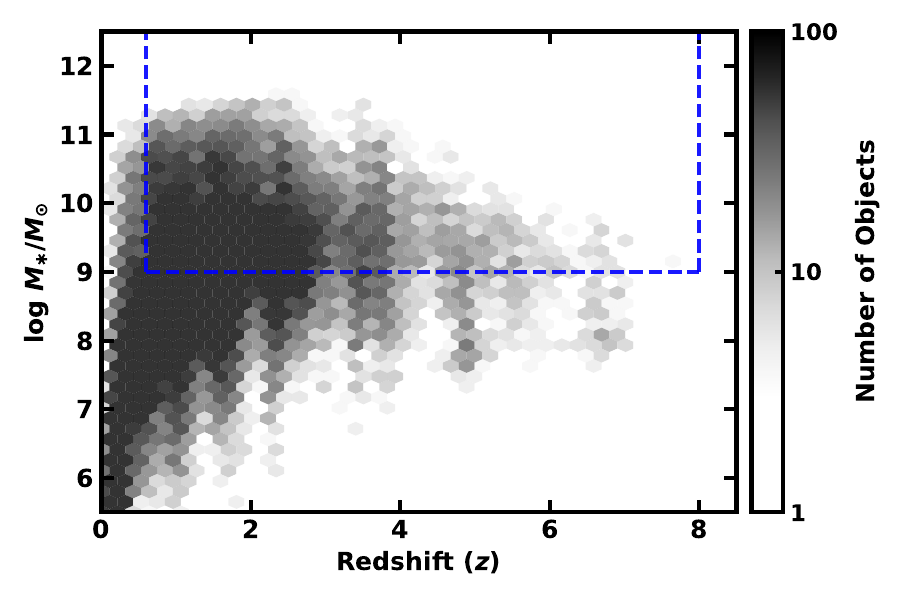}
\caption{
Distribution of stellar masses of extended sources in all JWST field as a function of redshift.
Stellar masses and redshifts are derived using the EAZY-PY model in this study.
The blue dashed lines represent the selection criteria of ${\rm log}~M_{\ast}/M_{\odot}>9$ and $z=0.6-8.0$, which are used to select the galaxy candidate for morphology classification.
\label{fig:mass_vs_z}}
\end{figure}

\subsubsection{Sample Selection Based on Redshifts and Stellar Masses}

This section describes our sample selection from the results of SED fitting.
Initially, we established a robust sample by excluding the sources with high chi-square values ($\chi^{2}_{\rm SED}\geq20$), smaller number of filters ($N_{\rm filt}<6$), and implausible output parameters such as negative values for photometric redshift, chi-square, stellar mass, and rest-frame fluxes.
In addition, we rejected faint objects with ${\rm F200W}>28~{\rm mag}$ from our sample due to their low signal-to-noise ratios, which could be inappropriate for the investigation of morphologies \citep{hue23, kar23}.
The numbers of these robust samples from the SED fitting are listed in the third column in {\color{blue} \textbf{Table \ref{tab:samp}}}.
We then selected the high-redshift galaxy samples for the morphological analysis based on redshifts and stellar masses.
{\color{blue} \textbf{Figure \ref{fig:mass_vs_z}}} illustrates the stellar mass distribution for the selected objects across all JWST fields, as a function of redshift.
We utilized spectroscopic redshifts when available, otherwise we employed photometric redshifts obtained through SED fitting.
Our final sample for studying morphologies was chosen with a redshift range from $z=0.6$ to $z=8.0$ and a stellar mass range of ${\rm log}~M_{\ast}/M_{\odot}>9$, as represented in the figure.
For the galaxies in the SMACS0723 field, we carefully selected a sample at the redshift range of $z=1.2-8.0$ to minimize any potential contamination from the foreground SMACS0723 cluster.
These selection criteria ensure that our sample is not significantly affected by completeness issues of observational data in the selected redshift range.
The stellar mass range chosen in this study was commonly employed in most JWST studies focusing on galaxy morphology \citep{fer23, hue23, kar23}.
The specific number of objects selected based on redshifts and stellar masses for each field is detailed in the fourth column in {\color{blue} \textbf{Table \ref{tab:samp}}}.

\begin{deluxetable*}{cccccc}
	\tabletypesize{\footnotesize}
	\setlength{\tabcolsep}{0.30in}
	\tablecaption{Numbers of Objects in Each Field for All Selection Steps}
	\tablehead{\colhead{Field} & \multicolumn{5}{c}{Number of objects for each step} \\
    & \colhead{(1)} & \colhead{(2)} & \colhead{(3)} & \colhead{(4)} & \colhead{(5)}}
	\startdata
    NEP-TDF & 12,222 & 4,929 & 2,314 & 799 & 752 \\
    NGDEEP & 17,184 & 6,276 & 1,971 & 557 & 507 \\
    CEERS & 86,484 & 44,956 & 13,652 & 5,293 & 5,111 \\
    COSMOS & 65,430 & 35,620 & 9,829 & 4,119 & 3,982 \\
    UDS & 89,178 & 54,101 & 21,729 & 8,793 & 8,412 \\
    SMACS0723 & 11,930 & 6,532 & 2,488 & 538 & 351 \\ \hline
    Total & 282,428 & 152,414 & 51,983 & 20,099 & 19,115
	\enddata
	\label{tab:samp}
	\tablenotetext{}{\textbf{Notes.} Columns represent the numbers of the following source categories: (1) The total number of sources detected by \textsc{SExtractor}; (2) Extended source candidates chosen for the \textsc{eazy-py} SED fitting, meeting the criteria based on colors and sizes as described in Section \ref{sec:phot}; 
    (3) Robust samples obtained from the \textsc{eazy-py} SED fitting, based on criteria such as $0<\chi^{2}_{\rm SED}<20$, $N_{\rm filt}\geq6$, ${\rm F200W}<28~{\rm mag}$, and positive values for photometric redshifts, stellar masses, and rest-frame fluxes.
    These sources are depicted in {\color{blue} \textbf{Figure \ref{fig:mass_vs_z}}}; (4) Galaxy candidates with ${\rm log}~M_{\ast}/M_{\odot}>9$ at $z=0.6-8.0$; (5) The final sample selected through visual inspections and excluding non-converged sources with $n\leq\delta n$.
    We examine the morphologies of this final sample using \textsc{GALFIT}, as described in Section \ref{sec:result}.} 
\end{deluxetable*}

\section{Morphology Classification}
\label{sec:morp}

\subsection{Morphology Classification Scheme in the HR5 Study}
\label{sec:class}

For fair comparisons of the results on galaxy morphology among different studies, the classification scheme should be as close as possible.
To ensure consistency with the HR5 study, we adopted the same morphology classification scheme described in \citet{park22}.
In their work, approximately $33,000$ simulated galaxies with $M_{\ast}>2\times10^{9}~M_{\odot}$ were categorized into three distinct types: `disks', `spheroids', and `irregulars'.
As depicted in Figure 2 of \citet{park22}, their classification scheme relies on two key parameters: the S\'ersic index ($n$) and the asymmetry factor ($A$).
These parameters were derived from the spatial distribution of stellar particles belonging to each galaxy.

In \citet{park22}, 
symmetric galaxies were categorized with $A<0.4$, while asymmetric galaxies were characterized by $A>0.4$.
This criterion is similar to thresholds used in previous observational studies \citep{con03, fer23, kar23}, which differentiated undisturbed galaxies with $A<0.35$ and merging galaxies with $A>0.35$.
However, the asymmetry measurement in \citet{park22} was based on the three-dimensional distribution of stellar particles, which could lead to a systematic difference from the asymmetry calculated using two-dimensional observational data.
In this study, we adopted a criterion of $A=0.32$ derived from the direct comparison between three- and two-dimensional asymmetry parameters (refer to Appendix B for details).

In the HR5 study, 
the radial stellar mass distribution 
within galaxies was fit by the S\'ersic model.
Symmetric galaxies were further divided into disks disks ($n<1.5$) and spheroids ($n>1.5$) based on the results of S\'ersic profile fitting.
The choice of the S\'ersic index threshold of $n=1.5$ was based on the fact that it effectively distinguishes between late-type and early-type galaxies, as elaborated in the Appendix of \citet{park22}.
It is very important to note that the criterion $n=1.5$ results in the most accurate classification when the central part of galaxies is excluded in the S\'ersic profile fitting as this can remove the galaxy bulge component and any observational artifact/AGN near the center. When the central part is included in the fitting, disk galaxies with a bright central bulge or core star-burst will become difficult to be distinguished from spheroidal ones. 

Irregular galaxies were categorized to include asymmetric galaxies ($A>0.32$) or galaxies where the S\'ersic fitting failed.
In the subsequent sections, we explain how we applied this classification scheme to categorize galaxy morphologies using JWST observational data.

\subsection{Measuring the Asymmetry}
\label{sec:asym}

The rotational asymmetry index has commonly used for quantifying the degree of asymmetry in galaxy light, as introduced by \citet{abr96}.
This index is determined by calculating the sum of differences between the original image and its corresponding $180^{\circ}$-rotated image, relative to the sum of the original image.
In \citet{park22}, the asymmetry of simulated galaxies was measured using the three-dimensional stellar mass density within pixels containing at least 10 stellar particles, excluding the central region within a radius of $0.8~{\rm kpc}$ in proper distance.

In this work, we followed a similar methodology to \citet{park22} with the additional correction for observational background noise, as recommended in \citet{con00} and \citet{lot04}.
The asymmetry index ($A$) was computed using the following formula,
\begin{equation}
    A = \frac{\Sigma_{i,j\in S}~|I_{0}(i,j)-I_{180}(i,j)|}{2\times\Sigma_{i,j\in S}~I_{0}(i,j)}-A_{\rm bkg},
    \label{eqn:asym}
\end{equation}
where $I_{0}$ and $I_{180}$ are the pixel values in the original and $180^{\circ}$-rotated images, and $A_{\rm bkg}$ is the asymmetry index for the background region.
The indices of $i$ and $j$ denote the pixels within the selected region ($S$) used to measure asymmetry.

To apply the formula, we followed these steps.
Initially, we created cutout images for all selected galaxies with a size of $10''\times10''$, which is sufficient to cover the total fluxes of the galaxies.
The filter selection for these cutout images was tailored to the redshift of each galaxy, as illustrated in {\color{blue} \textbf{Figure \ref{fig:wav}}}.
Then, we ran \textsc{SExtractor} twice on the cutout images, with a reduced background mesh size of 32, and keeping other input configurations consistent with those listed in {\color{blue} \textbf{Table \ref{tab:sep}}}.
The segmentation map from the initial \textsc{SExtractor} run was used to mask any neighboring light sources adjacent to the target galaxy.
This masking process was implemented by replacing the pixel values of the detected contaminants with the background value of the image, creating a masked cutout image.
In the second \textsc{SExtractor} run with the masked cutout image, we were able to more precisely measure the Petrosian radius \citep[$r_{p};$][]{pet76} of each galaxy.
We also generated a local background image around the galaxy within the masked image, by utilizing a two-dimensional linear function to fit any possible background gradients.
Although the obtained JWST images were already background-subtracted, these cutout images might still contain subtle background gradients.
This background gradient model was used in the subsequent \textsc{GALFIT} analysis described in Section \ref{sec:galfit}.

To calculate the asymmetry parameter, we selected a specific region for the detected source ($S$) within an elliptical aperture with a semi-major axis of $1.5\times r_{p}$ from the galaxy center.
The radius of $1.5\times r_{p}$ is known to encompass nearly the total flux of a galaxy \citep{con00, lot04}.
The inner circular region with a radius of $0.8~{\rm kpc}$ was excluded for this calculation, as done in \citet{park22}.
The pixel values within the region $S$ were summed without any smoothing procedures.
To correct for the background noise, we designated the background region as the area outside a radius of $1.5\times r_{p}$ from the center of the galaxy.
For determining $A_{\rm bkg}$, we adopted the minimum asymmetry value from the background regions in the $90^{\circ}$-, $180^{\circ}$-, and $270^{\circ}$-rotated images.
This minimum asymmetry value was then normalized by the area of the object region ($S$) when applying Equation \ref{eqn:asym}.

\subsection{Surface Photometry with GALFIT}
\label{sec:galfit}

We utilized \textsc{GALFIT} \citep{pen02, pen10} to analyze the morphology of galaxies based on their light distribution.
We used the masked cutout image of each galaxy, as mentioned in Section \ref{sec:asym}, as the input image for running \textsc{GALFIT}.
We applied a single S\'ersic profile with free parameters for the central coordinates, total magnitude, effective radius, and S\'ersic index.
The axis ratio and position angle were fixed to the output values derived from the second \textsc{SExtractor} run described in Section \ref{sec:asym}, in order to enhance the stability of \textsc{GALFIT} solutions and speed up the fitting procedure.
In addition, we applied a two-dimensional sky component with a free sky background value and fixed sky gradients along the x and y axes.

It is important to note certain limitations in our choice of a single S\'ersic profile for \textsc{GALFIT} in this study.
Galaxies are known to exhibit multiple morphological components such as disks, bars, spiral arms, bulges, and irregular features resulting from merger or tidal processes.
In particular, around $10\%$ of galaxies have both bulge components ($n\sim4$) at their centers and exponential disk components ($n\sim1$) in their outskirts simultaneously \citep{hue23, kar23}.
This inherent complexity in galaxy morphology has led previous studies on high-redshift galaxy morphology to rely on visual classification and use parametric measurements from \textsc{GALFIT} as a supplement.
Ideally, the inclusion of multiple S\'ersic profiles in the \textsc{GALFIT} configurations would be more beneficial for tracing the detailed morphological structures of real galaxies.
However, multiple S\'ersic fitting does not work effectively for most high-redshift galaxies with small apparent size ($R_{\rm eff}\lesssim0\farcs2$), and it is very challenging to apply to a large sample of galaxies.
Therefore, this work opted to use a single S\'ersic profile for the consistency in methodology with the HR5 study and the convenience of analyzing a large sample of high-redshift galaxies.

In our \textsc{GALFIT} configurations, we employed point spread function (PSF) images generated from \textsc{WebbPSF} \citep{per14} to account for the PSF effect.
We resampled the PSF images of NIRCam bands to the pixel scale of $0\farcs04~{\rm pixel^{-1}}$, with the image sizes of $5\arcsec\times5\arcsec$.
The size of the PSF convolution box was set to be larger than the total area of the input images.
Since \textsc{WebbPSF} only creates the simulated PSFs of JWST/NIRCam, it is necessary to test the effect of empirical PSF models which can reflect the quality of real drizzled images \citep{ono23, zhu24}.
For this reason, we also generated empirical PSFs utilizing \textsc{PSFEx} \citep{ber11} and applied them to the \textsc{GALFIT} analysis of the whole sample.
We found that the choice of PSF models have little influence on our main results in Section \ref{sec:morp_Mz}.
We described the details of the tests for empirical PSFs in Appendix C.

To minimize the cases of obtaining unreasonable results, we imposed constraints on the S\'ersic index, ranging from 0.2 to 10, and the coordinate offsets to $0\farcs1$ (2.5 pixels) from the galaxy center. 
After executing \textsc{GALFIT}, 
we systematically rejected sources with unreliable solutions, 
with 
the uncertainty of the S\'ersic index ($\delta n$) exceeding the value of the S\'ersic index itself ($n\leq\delta n$).
The final sample size with this rejection is detailed in the fifth column of {\color{blue} \textbf{Table \ref{tab:samp}}}.

We 
evaluated the quality of the GALFIT fitting by computing the residual flux fraction \citep[$RFF$;][]{hoy11} for each galaxy.
The $RFF$ measures the signal in a residual image relative to the sum of the original image.
As in \citet{hoy12}, the $RFF$ is defined as,
\begin{equation}
    RFF=\frac{\Sigma_{i,j\in S}~|I(i,j)-I(i,j)^{\rm model}|-0.8\times\Sigma_{i,j\in S}~\sigma_{\rm B}(i,j)}{\Sigma_{i,j\in S}~I(i,j)}
    \label{eqn:rff}
\end{equation}
where $|I(i,j)-I(i,j)^{\rm model}|$ is the residual signal between the input image and the \textsc{GALFIT} model image.
$\sigma_{\rm B}$ denotes the background fluctuation, and $S$ is the region selected for this calculation.
In this study, we selected $S$ as an elliptical region within a semi-major axis of $1.5\times r_{p}$, excluding the central region within a radius of 0.8 kpc.
In terms of the background fluctuation, we adopted the background sigma for the total area of the masked cutout image, as assumed in \citet{mar16}.
The equation used for calculating the background fluctuation is as follows,
\begin{equation}
    \Sigma_{i,j\in S}~\sigma_{B}(i,j)=N_{S}\times\langle\sigma_{\rm B}\rangle,
    \label{eqn:sigma}
\end{equation}
where $N_{S}$ is the number of pixels belonging to the area used for $RFF$ calculation, and $\langle\sigma_{B}\rangle$ is the mean background sigma in the image.
We determined $\langle\sigma_{B}\rangle$ to be the clipped sigma value from the region outside the Petrosian aperture ($S$).

We utilized the $RFF$ as a proxy to assess the success of a single S\'ersic fitting.
The specific $RFF$ criterion has not been agreed well because it can vary depending on the quality of observational data and the scientific purposes of using the $RFF$.
In a recent JWST study by \citet{orm23} and \citet{war23}, the $RFF$ was employed as a criterion for excluding poorly-fit objects with $RFF>0.5$.
This threshold was found to be more generous than previous HST studies, which suggested that objects with $RFF>0.11$ are needed to fit with additional S\'ersic components \citep{hoy11}, or merger candidates could be effectively found with $RFF>0.2$ (symmetric cases) or $RFF>0.1$ (asymmetric cases).
In this work, we adopted the criterion of $RFF>0.5$ to identify obviously poorly-fit objects from the GALFIT analysis along with avoiding any misclassification of multi-S\'ersic objects as poorly-fit ones.

\subsection{Parametric Morphological Classification in This Study}
\label{sec:mclass}

\begin{figure*}
\centering
\includegraphics[width=0.90\textwidth]{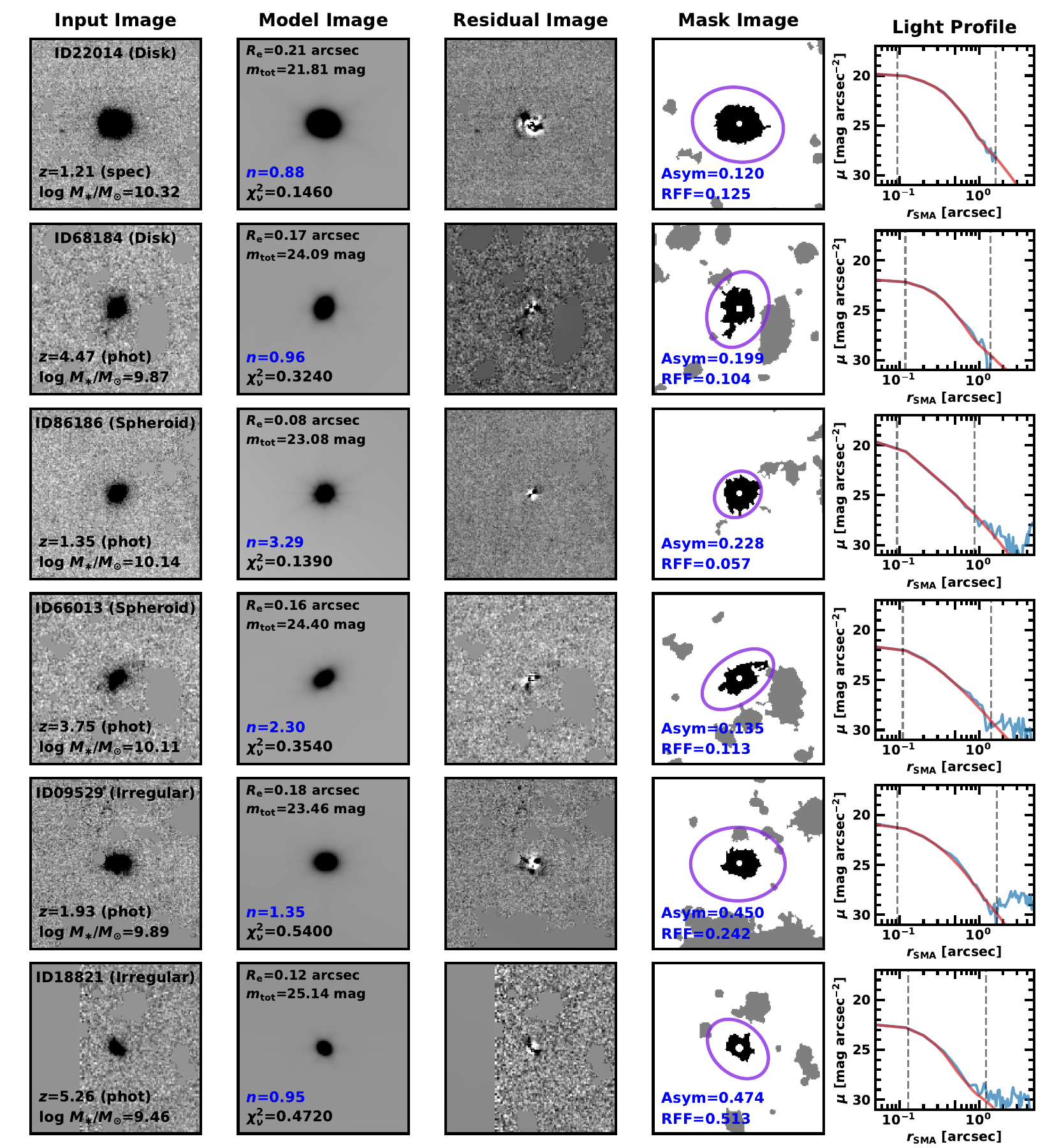}
\caption{
Thumbnail images ($6''\times6''$) of high-redshift galaxies selected in this study.
The examples include six galaxies, with two galaxies in each morphological category: disks (upper two objects), spheroids (middle two objects), and irregulars (lower two objects).
Each galaxy example is represented by five panels.
The leftmost panels display the masked cutout images used as input for \textsc{GALFIT}.
The subsequent two panels display the model images and residual images obtained from \textsc{GALFIT}.
In the fourth panels, the mask images show the distribution of masked pixels in the gray region, with the black region representing the $1\sigma$ boundary of light from objects recorded in the segmentation map.
Purple ellipses represent apertures with a semi-major axis of $1.5\times r_{p}$ of each galaxy.
The fifth panels show the radial light profiles measured from the galaxy center to the image edge.
Original light profiles are depicted by blue curves, while the model light profiles from \textsc{GALFIT} are depicted by red curves.
The two gray dashed lines represent a semi-major axis corresponding to 0.8 kpc (left) and a semi-major axis of $1.5\times r_{p}$ of the galaxy (right).
Key parameters for morphology classification are denoted by blue texts in this figure.
\label{fig:galfits}}
\end{figure*}

Utilizing the results obtained from asymmetry measurements and \textsc{GALFIT} fitting, we applied the following parametric criteria for morphology classification:

(1) Disk-type galaxies: $n<1.5$, $A<0.32$, and $RFF<0.5$,

(2) Spheroid-type galaxies: $n>1.5$, $A<0.32$, and $RFF<0.5$,

(3) Irregular-type galaxies: $A>0.32$ or $RFF>0.5$.\\
Symmetric and well-fitted galaxies with $A<0.32$ and $RFF<0.5$ were further categorized into disk-type ($n<1.5$) and spheroid-type galaxies ($n>1.5$) based on their S\'ersic indices.
Irregular-type galaxies in our analysis encompass both asymmetric galaxies with $A>0.32$ and poorly-fit galaxies with $RFF>0.5$.
This parametric classification scheme is consistent with the criteria applied in \citet{park22}.

{\color{blue} \textbf{Figure \ref{fig:galfits}}} illustrates the examples of our morphological classification.
Disk galaxies in the top two panels show nearly exponential light profiles ($n\sim1$), occasionally accompanied by spiral arm structures (ID 22014).
Spheroid galaxies in the middle two panels are characterized by bulge-like structures with higher S\'ersic indices ($n\gtrsim2$) compared to disk galaxies.
In the bottom two panels, irregular galaxies exhibit more remarkable substructures in their residual images compared to disks and spheroids.
Although there could be a potential problem of misjudgement near the sharp classifying boundary of parameters, our parametric classification scheme generally aligns with the common visual classification of galaxy morphology for the majority of galaxies in our extensive sample.

\section{Results}
\label{sec:result}

\subsection{Distributions of Asymmetry, $RFF$, and S\'ersic indices}
\label{sec:A_RFF}

\begin{figure*}
\centering
\includegraphics[width=0.85\textwidth]{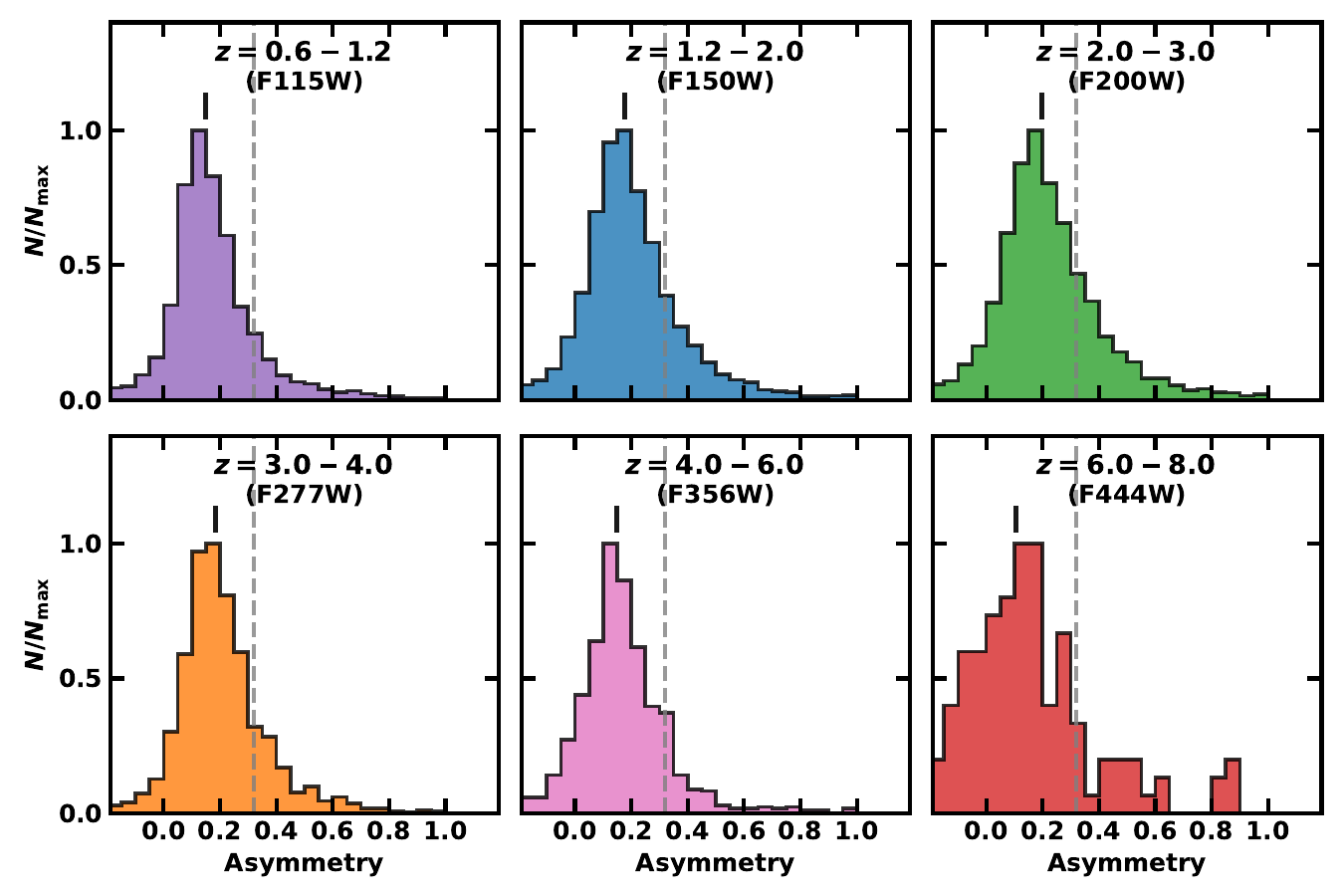}
\caption{
Histograms depicting the asymmetry of galaxies in all JWST fields.
The panels are organized based on the chosen JWST/NIRCam filters corresponding to the respective redshift ranges.
The gray dashed lines represent our criterion to distinguish symmetric galaxies ($A<0.32$) and asymmetric ($A>0.32$) galaxies.
The black lines mark the median values of asymmetry in each redshift bin.
\label{fig:Asym}}
\end{figure*}

\begin{figure*}
\centering
\includegraphics[width=0.85\textwidth]{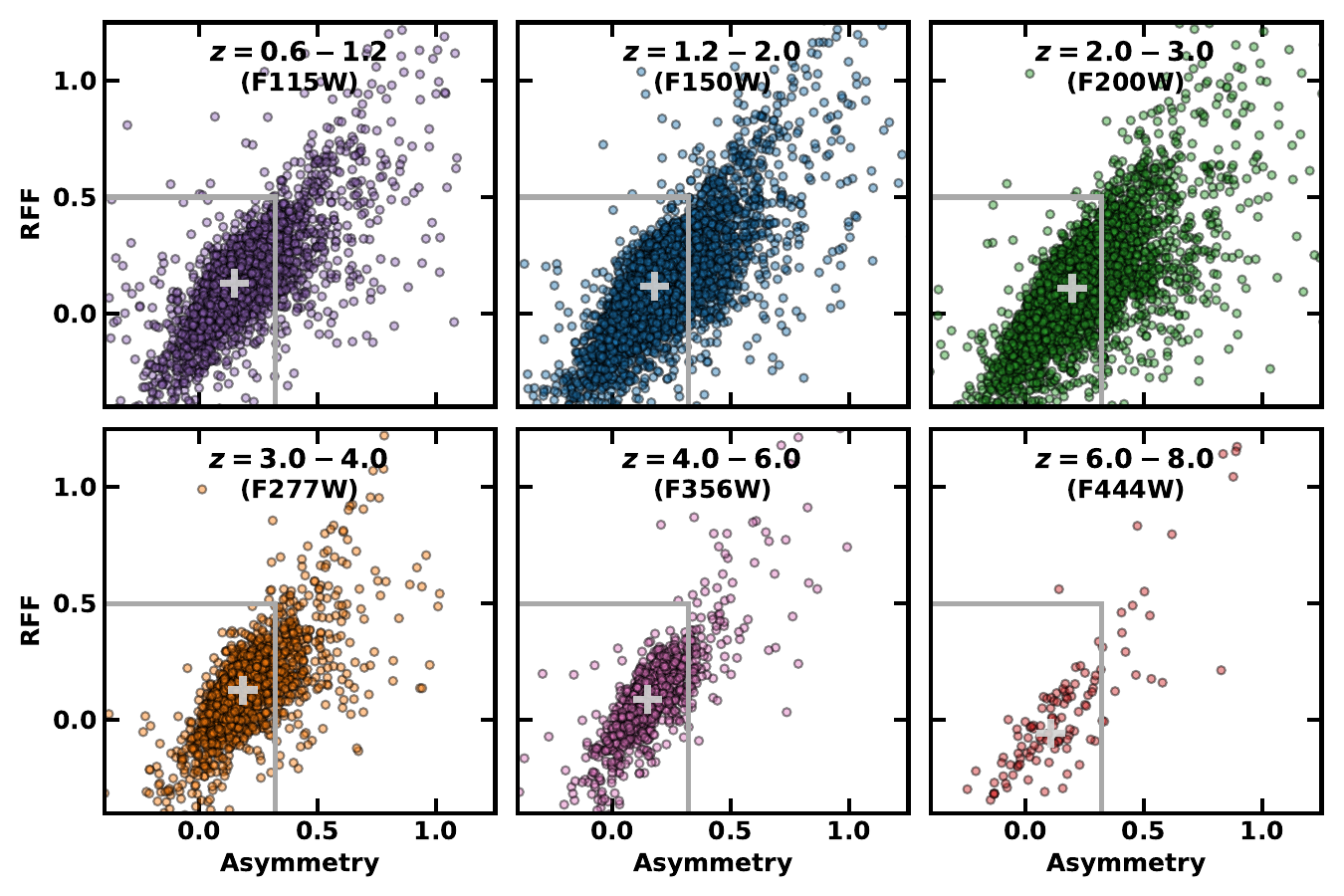}
\caption{
Distributions of asymmetry and $RFF$ values of our sample. 
The panels are ordered as in  {\color{blue} \textbf{Figure \ref{fig:Asym}}}.
The gray lines denote the selection criteria of $RFF=0.5$ and $A=0.32$ to divide regular galaxies (disks and spheroids) and irregular galaxies in this study.
The gray cross symbols mark the median values of $RFF$ and asymmetry in each redshift bin.
\label{fig:RFF_vs_Asym}}
\end{figure*}

\begin{deluxetable*}{ccccccc}
	\tabletypesize{\footnotesize}
	\setlength{\tabcolsep}{0.20in}
	\tablecaption{Statistics of asymmetry, $RFF$, and S\'ersic index of Galaxies in Each JWST Field}
	\tablehead{\colhead{Field} & \multicolumn{2}{c}{Asymmetry} & \multicolumn{2}{c}{$RFF$} & \multicolumn{2}{c}{S\'ersic index} \\
    & \colhead{Median} & \colhead{$\sigma_{\rm MAD}$} & \colhead{Median} & \colhead{$\sigma_{\rm MAD}$} & \colhead{Median} & \colhead{$\sigma_{\rm MAD}$}}
	\startdata
    NEP-TDF & 0.18 & 0.11 & 0.14 & 0.11 & 1.06 & 0.39 \\
    NGDEEP & 0.24 & 0.09 & 0.24 & 0.08 & 1.06 & 0.37 \\
    CEERS & 0.19 & 0.08 & 0.16 & 0.09 & 1.09 & 0.40 \\
    COSMOS & 0.17 & 0.09 & 0.12 & 0.09 & 1.09 & 0.41 \\
    UDS & 0.16 & 0.09 & 0.08 & 0.10 & 1.06 & 0.40 \\
    SMACS0723 & 0.20 & 0.10 & 0.16 & 0.12 & 1.03 & 0.37 \\ \hline
    Total & 0.18 & 0.09 & 0.12 & 0.10 & 1.08 & 0.40 \\
	\enddata
	\label{tab:A_rff}
\end{deluxetable*}

In this section, we present the distributions of key parameters for morphology classification: asymmetry, $RFF$, and S\'ersic indices.
{\color{blue} \textbf{Figure \ref{fig:Asym}}} shows the distributions of asymmetry values of our sample galaxies.
The asymmetry distributions exhibit a strong concentration around $A\sim0.18$ across all redshift ranges.
As also shown in {\color{blue} \textbf{Table 5}}, the median asymmetry values of galaxies in each JWST field are also around $A\sim0.18$, indicating no systematic biases in asymmetry measurements depending on filters and fields.
{\color{blue} \textbf{Figure \ref{fig:Asym}}} also shows elongated tails towards higher asymmetry values in all redshift ranges.
These tails are indicative of irregular galaxies with mergers and remarkable substructures.
As mentioned in Section \ref{sec:class}, we classified galaxies as asymmetric when $A>0.32$, which is consistent with the criterion of three-dimensional asymmetry with $A>0.4$ in \citet{park22}.

{\color{blue} \textbf{Figure \ref{fig:RFF_vs_Asym}}} displays the distributions of $RFF$ and asymmetry in each redshift range.
We note that there is a positive correlation between $RFF$ and asymmetry, implying that asymmetric galaxies have high residual fluxes from a single S\'ersic fitting.
The $RFF$ distributions consistently show median values of $RFF\sim0.12$ at $z<4$, whereas at $z>4$, the median $RFF$ values become lower than $0.1$.
This decrease in $RFF$ at higher redshifts might be due to the reduced effective radii in both physical and angular sizes of galaxies in the early universe.
In contrast to asymmetry, there are some systematic variations in $RFF$ values of galaxies depending on the JWST fields, as shown in {\color{blue} \textbf{Table 5}}.
These field-to-field variations result from differences in background fluctuations within the images of the JWST fields.
For instance, the NGDEEP field shows relatively higher $RFF$ with a median value of $0.24$ and $\sigma_{\rm MAD}=0.08$, compared to other fields with a median value of $0.12$ and $\sigma_{\rm MAD}=0.10$.
This is because the JWST images of NGDEEP have a deeper surface brightness limit, leading to lower background sigma in Equation \ref{eqn:rff}.
However, these systematic variations across JWST fields have a negligible effect when applying the generous criterion of $RFF>0.5$ to identify poorly-fit galaxies.
In addition, a majority of galaxies classified as irregulars are selected with the asymmetry criterion of $A>0.32$, so the $RFF>0.5$ criterion 
has an insignificant effect on the classification of irregular galaxies.

\begin{figure*}
\centering
\includegraphics[width=0.85\textwidth]{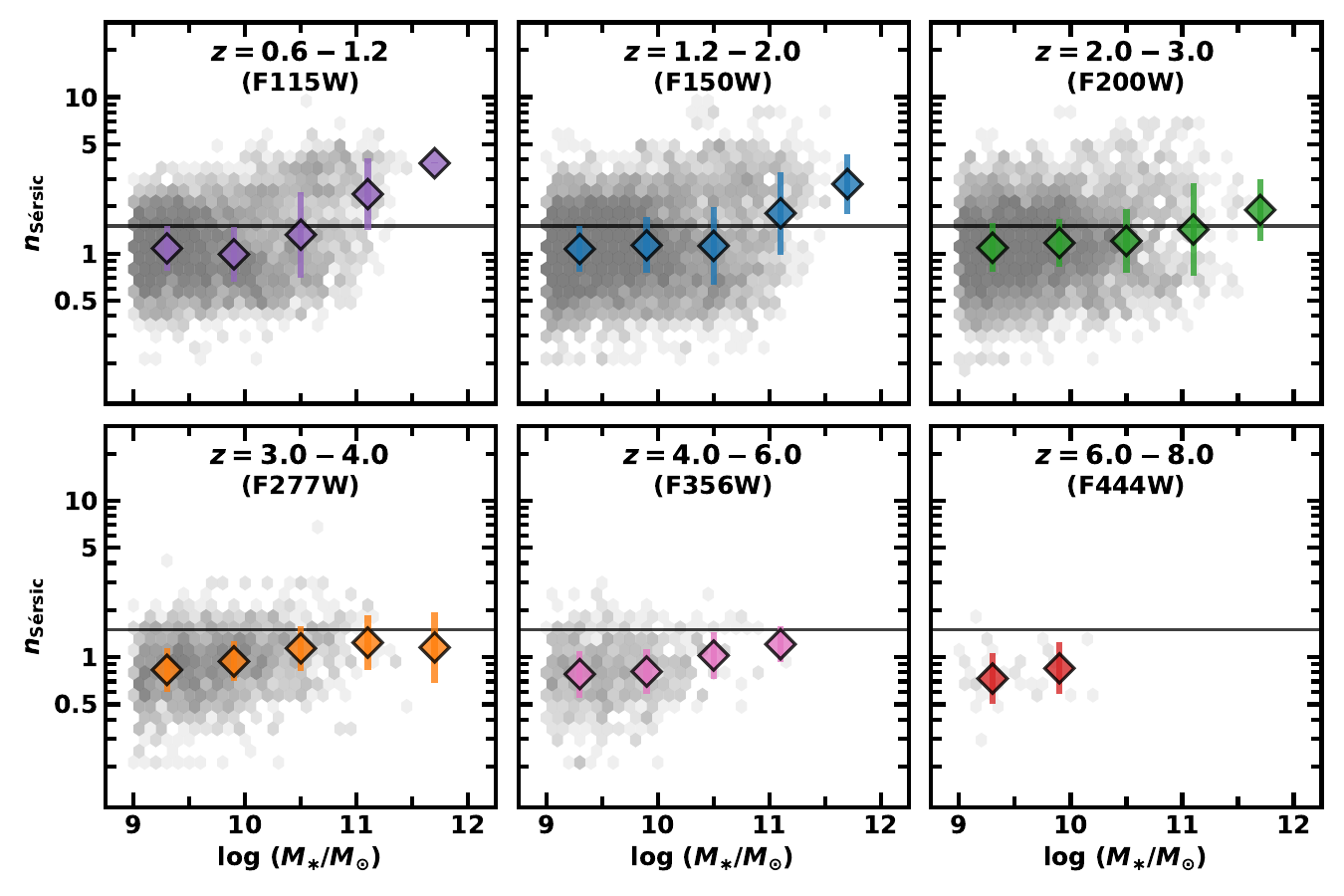}
\caption{
Distributions of S\'ersic indices as a function of stellar mass.
The panels are ordered as in {\color{blue} \textbf{Figure \ref{fig:Asym}}}.
The median values of S\'ersic indices in each stellar mass bin are represented by diamond symbols, along with their deviations.
The black dashed lines mark the criterion of $n=1.5$ to distinguish disk galaxies and spheroid galaxies.
In this figure, we do not plot median values for stellar mass bins with fewer than five objects.
\label{fig:n_vs_M}}
\end{figure*}

{\color{blue} \textbf{Figure \ref{fig:n_vs_M}}} displays the distribution of S\'ersic index as a function of stellar mass and redshift.
Although we applied the $n=1.5$ criterion to classify disk and spheroid galaxies, our analysis reveal continuous distributions of S\'ersic indices across all redshift ranges.
The reason of our choice of $n=1.5$ is described in Section \ref{sec:class} and the Appendix of \citet{park22}.
Overall, S\'ersic indices tend to increase in the high-mass regime of ${\rm log}~M_{\ast}/M_{\odot}\gtrsim10.5$, indicating that galaxy mass growth via mergers is accompanied by bulge-like structures with higher S\'ersic indices.
Examining the redshift evolution of S\'ersic indices, we find that galaxies at $z>3$ exhibit slightly lower S\'ersic index compared to those at $z<3$. 
With our classification framework, spheroid galaxies with $n>1.5$ appear more dominant than disk galaxies in the ${\rm log}~M_{\ast}/M_{\odot}\gtrsim10.5$ range at $z<3$ and become rare at $z>3$.
As shown in {\color{blue} \textbf{Table 5}}, the distribution of S\'ersic index is statistically consistent across 
all the JWST fields with a very small field-to-field fluctuation of the median $n$.



\subsection{Morphological Distribution with Stellar Mass and Redshift}
\label{sec:morp_Mz}

\begin{figure*}
\centering
\includegraphics[width=0.9\textwidth]{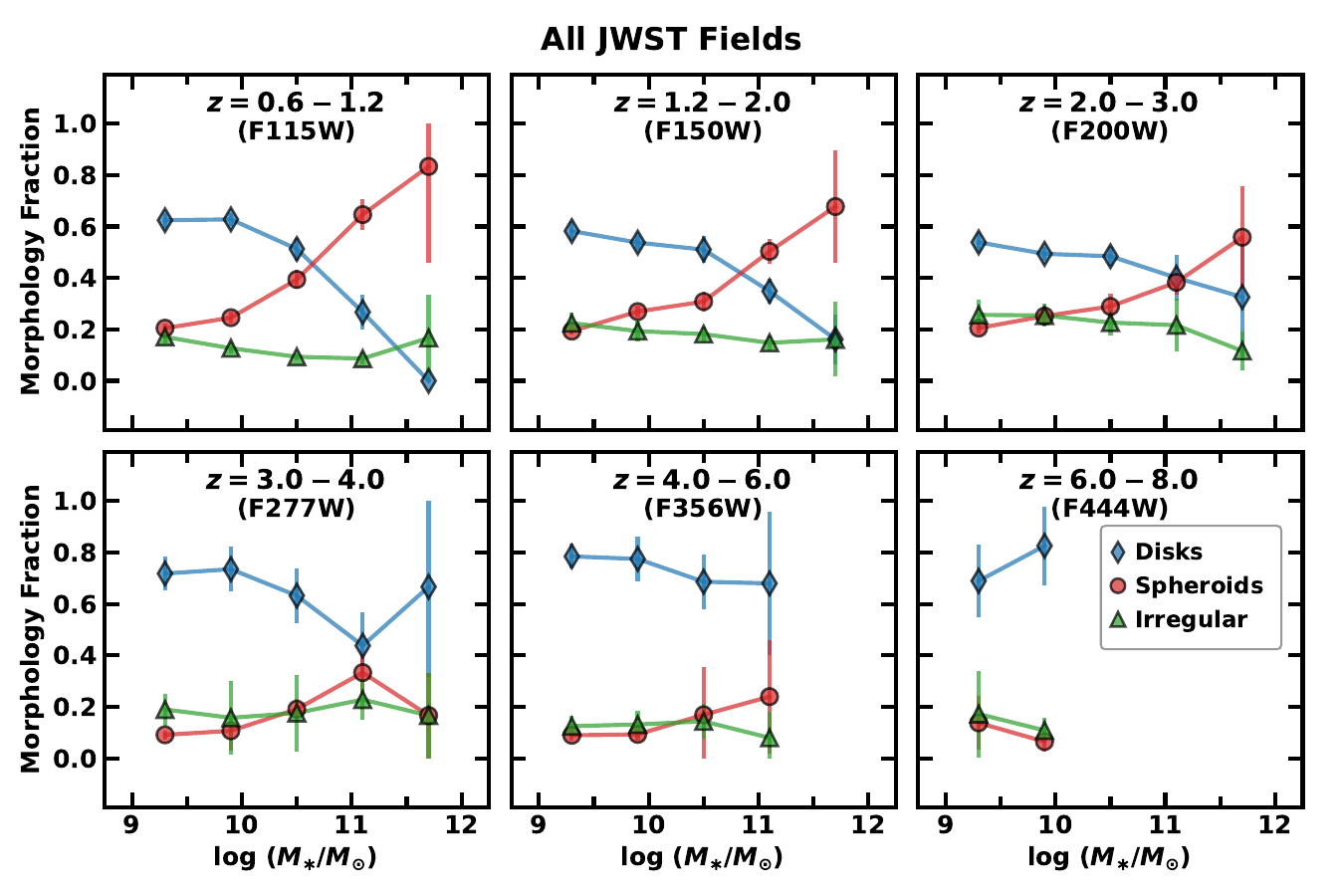}
\caption{
Morphological fractions as a function of stellar mass in various redshift bins.
The fractions of disks, spheroids, and irregulars are denoted by blue diamonds, red circles, and green triangles, respectively.
We plot the data from all six JWST fields.
We do not plot the fractions for stellar mass bins with fewer than five objects.
\label{fig:mfrac_Mz}}
\end{figure*}

\begin{figure*}
\centering
\includegraphics[width=0.9\textwidth]{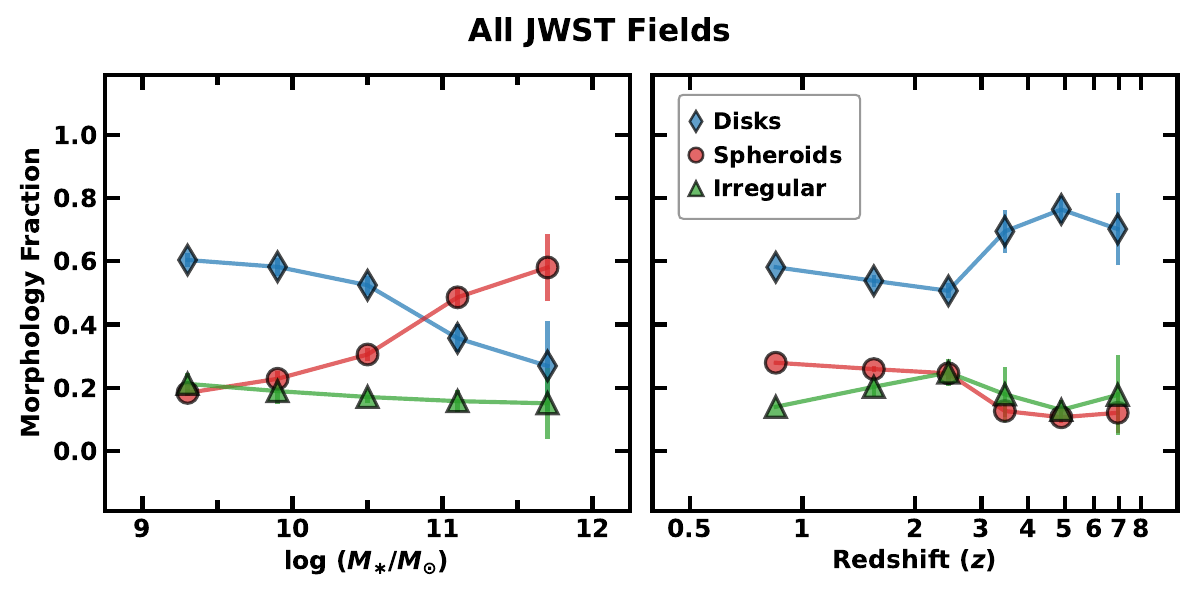}
\caption{
Morphological fractions as a function of stellar mass (left panel) and redshift (right panel).
Symbols are the same as {\color{blue} \textbf{Figure \ref{fig:mfrac_Mz}}}.
Here we also collect the data from all six JWST fields.
\label{fig:mfrac_Mztot}}
\end{figure*}

Here, we explain the morphological fractions in various stellar mass and redshift ranges, utilizing data collected from all six JWST fields.
It is worth noting that, unlike the other five blank fields, SMACS0723 has a dense galaxy cluster at $z=0.39$.
This may lead to significant interference from bright light emitted by cluster members and the intracluster medium.
In addition, the gravitational lensing effect induced by the SMACS0723 cluster can magnify and distort the morphologies of high-redshift galaxies.
Despite these potential problems, the general trends of morphological fractions from the small sample of the SMACS0723 field seem to be quite consistent with those in other fields (see Appendix A).
Thus, we use the data from the SMACS0723 field for {\color{blue} \textbf{Figures \ref{fig:mfrac_Mz}, \ref{fig:mfrac_Mztot}, and \ref{fig:mfrac_Mz3}}}, with only excluding galaxies at $z=0.6-1.2$.

{\color{blue} \textbf{Figure \ref{fig:mfrac_Mz}}} shows morphological fractions relative to stellar mass in redshift bins corresponding to different JWST filters.
To calculate uncertainties in each bin, we considered both Poisson noise from the sample size and field-to-field variations in morphological fractions.
Detailed information on morphological fractions for each JWST field is provided in Appendix A, along with the calculation method for uncertainties.
{\color{blue} \textbf{Figure \ref{fig:mfrac_Mz}}} reveals a clear dependence of disk and spheroid fractions on stellar mass.
In the ${\rm log}~M_{\ast}/M_{\odot}<10$ mass regime, disk galaxies are dominant constituting $\gtrsim60\%$ at all redshift ranges.
Spheroid fractions increase with stellar masses at all redshifts.
At $z<3$, spheroid galaxies become dominant in the high-mass regime, specifically ${\rm log}~M_{\ast}/M_{\odot}>10.5$ at $z<2$ and ${\rm log}~M_{\ast}/M_{\odot}>11$ at $z=2-3$.
Beyond $z>3$, spheroid galaxies consistently show lower proportions than disk galaxies across all stellar mass ranges.
Nevertheless, increasing trends in spheroid fractions still appear from the mass range of ${\rm log}~M_{\ast}/M_{\odot}\gtrsim10$.
In the case of irregular galaxies, their fractions remain nearly constant at around $20\%$ in all ranges of stellar mass and redshift.

We plot the combined data of morphological fractions in relation to stellar mass and redshift in {\color{blue} \textbf{Figure \ref{fig:mfrac_Mztot}}}.
The left panel shows the dependence of morphological fraction on stellar mass, combining data at all redshift ranges.
As stellar mass increases, the disk fraction decreases and the spheroid fraction increases; the crosspoint of disk and spheroid fractions occurs around ${\rm log}~M_{\ast}/M_{\odot}\sim10.5-11.0$.
The irregular fraction is nearly independent on stellar mass but slightly increases as stellar mass decreases in the range of ${\rm log}~M_{\ast}/M_{\odot}<11$, showing a larger fraction than spheroid fraction at ${\rm log}~M_{\ast}/M_{\odot}\lesssim10$.
The right panel shows the morphological fractions with redshifts, providing an insight on cosmic evolution of galaxy morphology.
In the early universe at $z>3$, disk galaxies are a little more prevalent than at $z<3$, with a fraction of $\gtrsim70\%$ compared to $\sim60\%$ at $z<3$,
while spheroid fraction increases at this cosmic period. This is due to the mass dependence of galaxy morphology and the fraction of massive galaxies increases at lower redshifts.
The spheroid fraction is only $\sim10\%$ at $z>3$, but reaches $\gtrsim20\%$ at $z<2$.
Irregular fraction does not show a monotonic relation with redshifts.
The irregular fraction seems to reach a maximum during the cosmic noon, $z=2.0 - 3.0$.

The dominance of disk galaxies at high redshifts and at relatively low stellar mass implies that the initial morphology of galaxies is disk-like when they first form.
Spheroidal and irregular galaxies also exist beyond $z>3$, but these galaxies constitute only about $10-20\%$. 
Changes in morphological fractions with stellar mass and redshift suggest that spheroid and irregular galaxies can be formed from initial disk galaxies via mergers or interactions.
In the redshift range of $z\lesssim2$, some of irregular galaxies seem to evolve into disk or spheroid galaxies, leading to a decrease in irregular fraction.

\section{Discussion}
\label{sec:discuss}

\subsection{Comparison with the HR5 Results}
\label{sec:HR5}

\begin{figure*}
\centering
\includegraphics[width=0.95\textwidth]{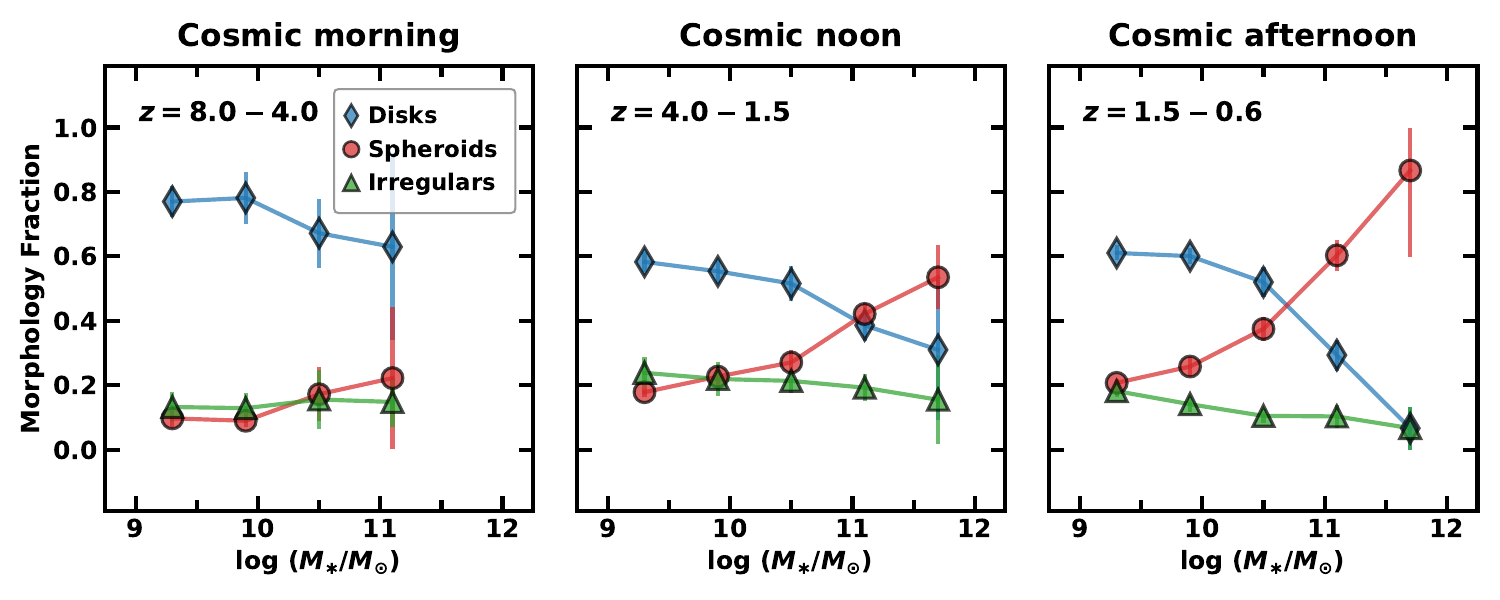}
\caption{
Morphological fractions as a function of stellar mass at cosmic afternoon ($z<1.5$; left panel), cosmic noon ($z=1.5-4.0$; middle panel), and cosmic morning ($z>4$; right panel).
Symbols are the same as {\color{blue} \textbf{Figure \ref{fig:mfrac_Mz}}}.
As in {\color{blue} \textbf{Figure \ref{fig:mfrac_Mz}}}, we do not plot the fractions for stellar mass bins with fewer than five objects.
\label{fig:mfrac_Mz3}}
\end{figure*}

Our results on the morphology fraction of high-redshift galaxies agree closely with those of  \citet{park22} in many ways. It should be noted that \citet{park22}'s prediction had been presented before any JWST results were announced. 
In the HR5 simulation, about 70\% of the first galaxies with $9.3<{\rm log}~M_{\ast}/M_{\odot}\lesssim 10$ 
are found to have disk morphology at $z>4$ (refer to their Figure 11).
The rests are divided into irregulars and spheroids more or less equally.
Spheroid and irregular fractions are nearly independent of redshift and stellar mass within the mass range, 
each type constituting around $\sim1/6$ of the first galaxies.
For comparison with these HR5 results, we depict the morphological fractions with stellar mass in {\color{blue} \textbf{Figure \ref{fig:mfrac_Mz3}}}, categorizing the cosmic epochs of `cosmic morning ($z>4$)', `cosmic noon ($z=1.5-4$)', and `cosmic afternoon ($z<1.5$)', following  \citet{park22}'s definition.

The left panel of {\color{blue} \textbf{Figure \ref{fig:mfrac_Mz3}}}, showing the morphological fraction of the three galaxy types in the cosmic morning ($z>4$),  illustrates a stunning agreement between our JWST observational results and HR5 cosmological simulation predictions. 
This panel shows that disk galaxies account for about 
75\% of galaxies with ${\rm log}~M_{\ast}/M_{\odot}\lesssim10$ at $z>4$, while spheroid and irregular galaxies account for about $10\%$
and $15\%$, respectively.
These morphological distributions closely resemble those from the HR5 study, indicating that 
the theoretical predictions on the morphology of the first galaxies by HR5 are clearly supported by
our JWST results.

In one of the HR5 snapshots during the cosmic noon period ($z\sim2$), disk galaxies remain dominant, comprising around $\sim70\%$ of galaxies in the ${\rm log}~M_{\ast}/M_{\odot}\lesssim10$ range (refer to Figure 20 in \citet{park22}).
However, the disk fraction shows a sharp decline in the mass range ${\rm log}~M_{\ast}/M_{\odot}\gtrsim10.5$.
Meanwhile, the spheroid fraction surpasses disk fraction in the ${\rm log}~M_{\ast}/M_{\odot}>11$ range, and irregular fraction also reaches up to $\sim35\%$ at ${\rm log}~M_{\ast}/M_{\odot}\gtrsim10.5$.
These trends of disk and spheroid fractions in the HR5 simulation also appear in a very similar fashion in our JWST analysis, as demonstrated in the middle panel of {\color{blue} \textbf{Figure \ref{fig:mfrac_Mz3}}}.
The results from this study show disk dominance up to ${\rm log}~M_{\ast}/M_{\odot}\sim11$ and spheroid dominance at higher masses.
However, the irregular fraction in our observational study shows a somewhat different behavior from that in the HR5 simulation.
The fraction indeed reaches a maximum during the cosmic noon period, but is nearly constant with $\sim20\%$ regardless of stellar mass, in contrast to the jump at ${\rm log}~M_{\ast}/M_{\odot}=10-10.5$ found by \citet{park22}.
This discrepancy could arise from differences in irregular classifiers, such as asymmetry and $RFF$, which may not be matched perfectly with those used in the HR5 study.

Morphological fractions in the cosmic afternoon ($z<1.5$) in the right panel
of {\color{blue} \textbf{Figure \ref{fig:mfrac_Mz3}}}
provide an insight on the morphological evolution of galaxies from high redshifts ($z>4$) to $z\lesssim1$.
It seems that the disk fraction in the cosmic afternoon has not changed much since the cosmic noon.
On the other hand, the spheroid fraction shows a noticeable increase, reaching $\sim70\%$ in the high-mass regime (${\rm log}~M_{\ast}/M_{\odot}>11$).
The irregular fraction in the cosmic afternoon exhibits a gentle decline with increasing stellar mass, in contrast to the constant fraction of $\sim20\%$ in the cosmic noon.
These trends in spheroid and irregular galaxies suggest that massive irregular galaxies in the cosmic noon or morning are likely to evolve into spheroid galaxies as they kinematically settle down and maintain the spheroidal morphology.

In summary, we note the following points from the perspective of these three cosmic periods.
First, disk galaxies are dominant in the cosmic morning with the fraction of $\sim75\%$ for those with ${\rm log}~M_{\ast}/M_{\odot}\lesssim10$, implying that disks are the initial morphology of the first galaxies.
Second, morphological transformations from disks to spheroids or irregulars occur across all the cosmic epochs, 
particularly in the mass range of ${\rm log}~M_{\ast}/M_{\odot}\gtrsim10.5$.
Third, spheroid galaxies exist in the cosmic morning at relatively low masses, 
but dominates the galaxy population in the cosmic noon and afternoon in the high-mass regime of ${\rm log}~M_{\ast}/M_{\odot}\gtrsim10.5-11$.
Fourth, the irregular fraction increases from the cosmic morning to the cosmic noon but decreases again in the cosmic afternoon due to the reduced number of massive irregular galaxies.
Lastly, these trends of morphological fractions can imply an evolutionary sequence from disks to spheroids or irregulars, with massive irregular galaxies eventually transforming into spheroids in the cosmic afternoon.
Overall, these findings align very well with 
the morphological evolution of galaxies suggested in the HR5 simulation.

\subsection{Formation and Evolution of Disks from the Early Universe}
\label{sec:disk}

In this study, we observed a prevalence of disk-type morphology for the galaxies with stellar masses ${\rm log}~M_{\ast}/M_{\odot}\lesssim10$ at $z>4$, which aligns with the HR5 results.
Interestingly, our findings from JWST images, capturing the distribution of stellar light, exhibit a good agreement with the HR5 simulation based on stellar mass distribution.
\citet{park22} explained that the emergence of disk-type morphologies in the early universe can be attributed to the initial angular momentum in protogalactic clouds.
The initial angular momentum is gained from the inflow of cold gas into protogalaxies along primordial large-scale structures, which is consistent with the basis of the tidal torque theory \citep{pee69, whi84}.
According to this theory, the tidal field on protogalactic regions and corresponding velocity field are governed by the large-scale dark matter distribution, which keep galaxies to acquire angular momentum set up by the initial conditions. The competition between the tidal torque driving galaxies into disk type and mergers driving galaxies into irregulars and spheroids determines the morphology distribution. 
The dominance of disks in the cosmic morning can be elucidated with these underlying physical processes.

In {\color{blue} \textbf{Figure \ref{fig:mfrac_Mz3}}}, our analysis reveals $\sim20\%$ decrease in disk fraction at small masses during the cosmic noon and afternoon compared to the cosmic morning, in contrast to a notable increase in the spheroid fraction in the high-mass regime.
In the low-mass regime (${\rm log}~M_{\ast}/M_{\odot}<10$), there is a slight increase of $5-10\%$ in the irregular fraction from the cosmic morning to the cosmic noon, indicating a tendency for disk galaxies in this mass range to transform into irregular galaxies.

However, it is important to note that mergers do not inevitably lead to irreversible morphological changes from disks to other types.
Despite undergoing morphological transformations due to mergers or destructive interactions, the initial disk structures can be restored through the acquisition of angular momentum via gas accretion from the vorticity-rich surrounding matter \citep{wel14, park22}.
Indeed, the HR5 simulation supported this recovery process, demonstrating morphological fluctuations in initial disk galaxies with changing S\'ersic indices \citep[refer to Figure 12 in ][]{park22}.
Consequently, this disk recovery mechanism can effectively explain why the disk fraction remains high in the cosmic morning in spite of high merger rate at corresponding redshifts \citep{dun19}.

\subsection{Comparison with Other Observational Studies on Galaxy Morphology}
\label{sec:jwst}

For comparison with other observational studies, it is important to note that the morphology classification schemes employed in other studies are not consistent with ours. 
We applied the parametric classification described in Section \ref{sec:mclass} for consistent comparison with the HR5 study, but previous studies have used different morphology classification approaches with various scientific objectives.
Commonly employed methods are visual classification with measurements of CAS parameters or machine learning techniques applied to large datasets.
Thus, the direct comparison with other studies is difficult.
Furthermore, our parametric classification has a potential problem of misclassifying the galaxies with parameters around $n\sim1.5$ or $A\sim0.32$ as our classification adopts sharp boundaries.
Despite these systematic differences and limitations, we discuss both agreements and disagreements on morphological distribution and evolution with the results from previous studies.

Most studies using HST data disagree with our findings of disk dominance in the early universe.
Previous HST studies have suggested a prevalence of irregular galaxies with disturbed structures at $z\gtrsim4$, rather than the disk dominance \citep{pap05, cam11, mor13, hue16}.
Otherwise, a few studies have demonstrated high spheroid fractions of $\sim30\%$ at $z\gtrsim2$ in the rest-frame UV wavelength \citep{lot06, dah07}.
These discrepancies with our results may arise from the limited observational performance of HST for high-redshift galaxy morphology and differences of observing wavelength ranges in the rest frame.
This is also pointed out by several JWST studies, which suggested that results based on JWST images exhibit a higher incidence of regular morphologies, such as disks and spheroids, compared to HST \citep{fer22, jac23}.
Furthermore, JWST images are useful for uncovering hidden disk galaxies within the noise of HST images \citep{nel23, rob23}.

Recent JWST studies agree with our findings regarding disk dominance in the early universe.
\citet{fer22,fer23}, utilizing visual classification with JWST fields of the SMACS0723 and CEERS, reported that disk galaxies occupy $40-60\%$ of the entire galaxy population at $z=1.5-6.5$.
They found that the disk fraction increases by a factor of 10 when using JWST images compared to investigations based solely on HST images.
The spheroid and irregular fractions are nearly constant at $10-20\%$, $30-40\%$, respectively, with respect to redshifts.
\citet{kar23}, also employing visual classification in the CEERS field, presented a dominant disk fraction of $\sim60\%$ at $z\sim3$, declining to $\sim30\%$ beyond $z>6$.
Similar to our results, \citet{kar23} noted an increase in spheroid fraction in the high-mass regime (${\rm log}~M_{\ast}/M_{\odot}\gtrsim10.5$) across all redshift ranges.
They reported pure spheroid and irregular fractions of $\sim20\%$ at $z>3$, with a substantial number of galaxies exhibiting mixed morphological structures due to their classification scheme with seven morphological types.
\citet{hue23} and \citet{toh23} also utilized more detailed subdivisions in their classification scheme than ours, employing CNN and unsupervised machine learning techniques.
They found that irregular and clumpy galaxies become dominant beyond $z>3$, with high fractions of disk-like morphologies in ${\rm log}~M_{\ast}/M_{\odot}>11$.

On the other hand, there have been several concerns about potential systematic uncertainties in morphology classification using JWST/NIRCam data.
\citet{veg24} used a contrastive learning framework to investigate high-redshift galaxy morphology at $z\sim3-6$ in CEERS, calibrating with mock data from the TNG50-1 simulation.
Their results agreed with the disk dominance in the high-redshift universe, with approximately half of the galaxies classified as disks.
However, they also noted that some compact and prolate-shaped galaxies might have been misclassified as pure disk galaxies, resulting in a potential overestimation of the disk fraction at $z>3$.
This is also supported by \citet{pan24}, which insisted that the S\'ersic indices alone cannot distinguish prolate and oblate populations when considering the three-dimensional geometry of CEERS galaxies.
Also using the CEERS field, \citet{sun24} analyzed the structures of 347 galaxies at $z=4-9.5$, based on the two parameters of S\'ersic indices and axis ratio.
They carefully computed the uncertainties of the parameters, considering the influence of the pixel scale when drizzling images, the PSF effects, and the cosmological effects depending on redshifts, such as surface brightness dimming, angular resolution, and sensitivity.
Their robust tests showed that the redshift has little impact on the S\'ersic index measurements up to at $z\lesssim6$, but higher redshift could lead to underestimation of S\'ersic indices for compact and spheroid-like ($n\gtrsim2$) galaxies (see Figure B2 in their Appendix).
Due to the systematic underestimation of the S\'ersic indices, disk fraction at $z>6$ could be overestimated by spheroid-like galaxies misclassified as disk-like galaxies.
Despite these uncertainties, \citet{sun24} also agreed high incidence of disk galaxies at $z>4$, providing the lower limit of disk fractions of $\sim45\%$ with the parametric criteria of $n<1.5$ and $b/a<0.6$.
We note that the systematic effect studied by \citet{sun24} would not significantly affect the morphological distributions of the JWST galaxies measured in this paper because only $\sim20\%$ of our sample at $z=4-8$ have redshift higher than 6 and almost $90\%$ of the galaxies at $z>6$ are not so compact, having S\'ersic indices less than 1.7 \citep[see the middle panel of Figure B2 in][]{sun24}.

In summary, previous JWST studies cannot be directly compared with one another as the classification criteria are all different. 
We adopt a simple but objective classification scheme that has been proven to be effective and can be directly compared to the HR5 cosmological simulation results. 
Although there are some differences and uncertainties in specific trends of morphological distributions, most JWST studies are in line with the disk dominance in the early universe.

\section{Summary}
\label{sec:summary}

In this study, we examine the rest-frame optical morphologies of high-redshift galaxies using the JWST/NIRCam images obtained from six JWST fields:  NEP-TDF, NGDEEP, CEERS, COSMOS, UDS, and SMACS0723.
We select $\sim19,000$ high-redshift galaxies with stellar masses of ${\rm log}~M_{\ast}/M_{\odot}>9$ and redshifts of $z=0.6-8.0$, derived from the SED fitting procedure.
We apply a parametric morphological classification scheme utilized in the HR5 simulation \citep{park22} and compare our findings with the HR5 results.
The key parameters for morphology classification (S\'ersic index, asymmetry, and $RFF$) were derived from the JWST/NIRCam images through \textsc{SExtractor} photometry and a single S\'ersic fitting with \textsc{GALFIT}.
Following the methodology of the HR5 study, our classification scheme categorizes three morphological types: disks ($n<1.5$, $A<0.32$, and $RFF<0.5$), spheroids ($n>1.5$, $A<0.32$, and $RFF<0.5$), and irregulars ($A>0.32$ or $RFF>0.5$).
From these analyses, our main results can be summarized as follows.

\begin{enumerate}
\item Our photometric redshift measurements from the SED fitting show an outlier fraction of $\sim6\%$ with $\sigma_{\rm NMAD}=0.009$ for objects with $\chi^{2}_{\rm SED}<20$ and $N_{\rm filt}\geq6$.
This accuracy is superior to the previous studies using HST-only data, highlighting that JWST photometric data is beneficial to analyze the properties of a large sample of high-redshift galaxies.
However, several outliers at $z_{\rm spec}<1$ exhibit overestimated photometric redshifts with $z_{\rm phot}>1$, potentially indicating low-redshift contaminants in our sample.
\item The distribution of asymmetry is strongly concentrated at $A=0.2$ but skewed toward higher asymmetry values across all redshift ranges.
This skewness occurs due to the existence of asymmetric galaxies with irregular substructures.
The $RFF$ values have positive correlations with asymmetry, with a median $RFF=0.13$.
The S\'ersic indices tend to increase with stellar mass, implying that the formation of bulge-like structures is accompanied with mass growth of galaxies.
The proportion of galaxies with bulge-like structures ($n>1.5$) declines at $z>3$ compared to the $z<3$ universe.
Potential biases of these parameters depending on the JWST fields have negligible effects on morphology classification.
\item Disk galaxies are dominant in the mass range of ${\rm log}~M_{\ast}/M_{\odot}<10.5$, with the disk fraction reaching up to $\sim80\%$ at $z>4$ in this mass range.
However, the disk fraction decreases with increasing stellar mass across all redshift ranges.
This implies that disks are likely to be the initial morphology of galaxies in the early universe and experience morphological transformations to spheroids or irregulars.
\item Spheroid galaxies are present at $z>6$ with a fraction of $\lesssim10\%$ of the galaxy population.
However, the spheroid fraction increases as stellar mass increases and redshift decreases, becoming the dominant type in ${\rm log}~M_{\ast}/M_{\odot}\gtrsim10.5-11$ at $z<3$.
This spheroid dominance in the high-mass regime implies that the mass growth is associated with morphological transformation from disks to spheroids.
\item Irregular galaxies maintain a relatively constant fraction of $\sim20\%$ with respect to stellar mass.
In the morphological fraction as a function of redshift, however, the irregular fraction exhibits a peak at $z\sim2-3$ and decreases again below $20\%$ in the $z<1$ universe.
This decrease in irregular fraction is mainly driven by a decrease in the number of massive irregulars (${\rm log}~M_{\ast}/M_{\odot}\gtrsim10.5$), implying that irregular galaxies eventually transform into spheroid galaxies in the cosmic afternoon ($z<1.5$).
\item The morphological distributions in the cosmic morning ($z>4$) are consistent with the HR5 simulation, confirming the disk dominance in the early universe.
As suggested in \citet{park22}, disk galaxies in the cosmic morning evolve into spheroids and irregulars through merger and gas accretion.
Massive irregular galaxies appear to be an intermediate phase in the morphological transition from disks to spheroids.
These evolutionary tracks of galaxy morphology align well with the HR5 study, thus demonstrating that this study effectively verify the HR5 results with the JWST observational data.
\end{enumerate}

The close agreement between our observational morphology distribution measurement and HR5 simulation results opens the possibility that the origin and evolution of galaxy morphology can be understood from cosmological simulations. It remains to be seen how future observational results with larger number of galaxies and improved redshift accuracy will compare with our measurement and also with the predictions of future cosmological simulations.

\begin{acknowledgments}
JHL was supported by the National Research Foundation of Korea (NRF) grant funded by the Korean government (MSIT) (Nos. 2022R1A4A3031306 and 2023R1A2C1006261). HSH acknowledges the support by the National Research Foundation of Korea (NRF) grant funded by the Korea government (MSIT) (No. 2021R1A2C1094577).
This work is based 
on observations made with the NASA/ESA/CSA James Webb Space Telescope. The data were obtained from the Mikulski Archive for Space Telescopes (MAST) at the Space Telescope Science Institute (STScI), which is operated by the Association of Universities for Research in Astronomy, Inc., under NASA contract NAS 5-03127 for JWST.
This research is based on observations made with the NASA/ESA Hubble Space Telescope obtained from STScI.
Some/all of the data presented in this paper were obtained from the MAST at STScI. The specific observations analyzed can be accessed via \dataset[https://doi.org/10.17909/b7cb-xc61]{https://doi.org/10.17909/b7cb-xc61} (NEP-TDF), 
\dataset[https://doi.org/10.17909/hh6k-9z63]{https://doi.org/10.17909/hh6k-9z63} (NGDEEP), 
\dataset[https://doi.org/10.17909/fna5-we11]{https://doi.org/10.17909/fna5-we11} (CEERS), 
\dataset[https://doi.org/10.17909/tdx4-0a35]{https://doi.org/10.17909/tdx4-0a35} (COSMOS), 
\dataset[https://doi.org/10.17909/8kp6-9223]{https://doi.org/10.17909/8kp6-9223} (UDS), 
and \dataset[https://doi.org/10.17909/8nfp-hr63]{https://doi.org/10.17909/8nfp-hr63} (SMACS0723).

\end{acknowledgments}

\software{Astropy \citep{ast13, ast18, ast22}, \textsc{eazy-py} \citep{bra08, bra23b}, \textsc{GALFIT} \citep{pen02, pen10},  Matplotlib \citep{hun07}, Numpy \citep{har20}, \textsc{PSFEx} \citep{ber11}, Scipy \citep{vir20}, \textsc{Source Extractor} \citep{ber96}, \textsc{WebbPSF} \citep{per14}}

\clearpage

\appendix

\section{Field-to-field Variations in Morphological Distributions}

\restartappendixnumbering

\begin{figure}
\centering
\includegraphics[width=0.9\textwidth]{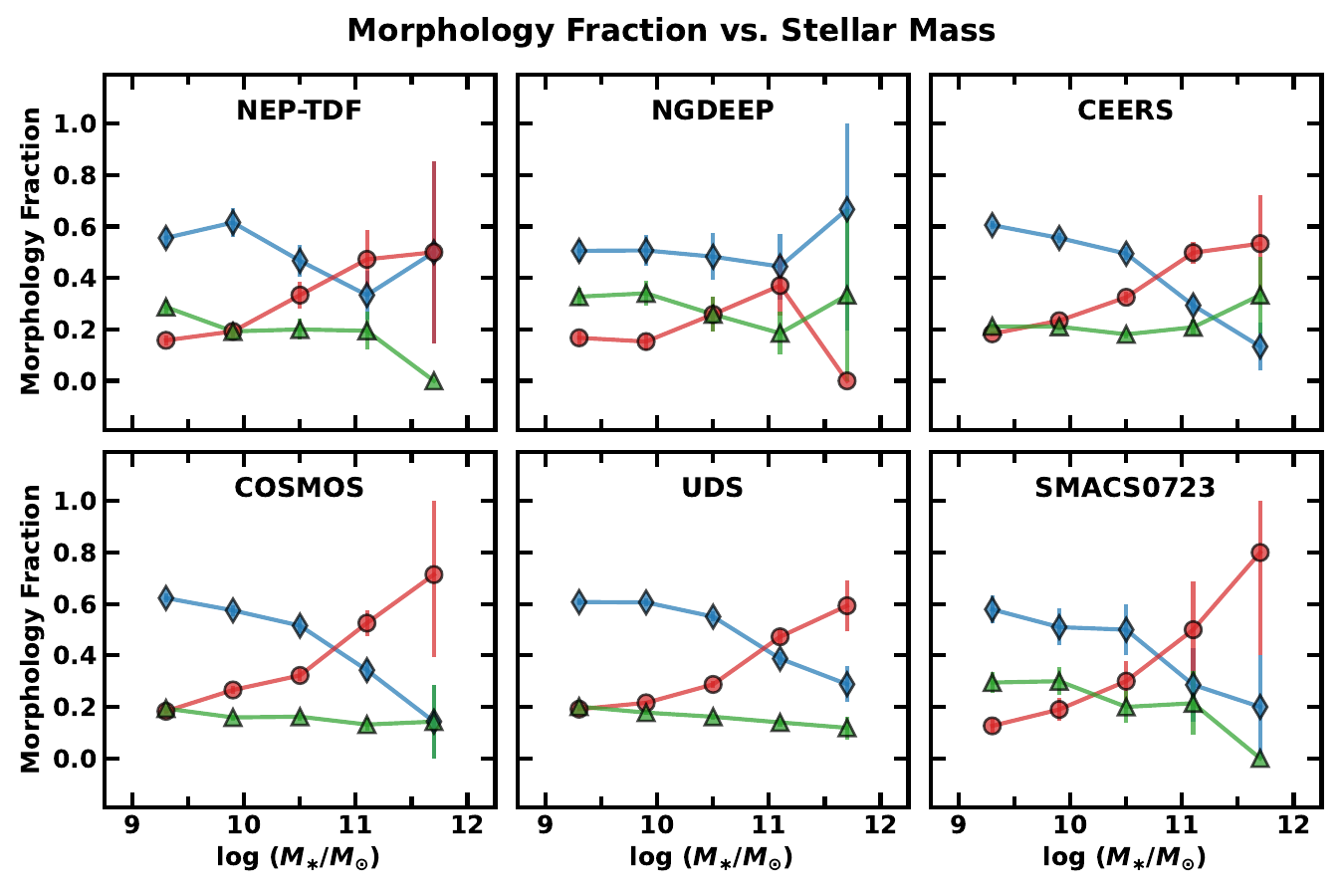}
\caption{
Same as the left panel of {\color{blue} {\bf Figure \ref{fig:mfrac_Mztot}}}, but for each JWST field.
\label{fig:apdx_mfrac_mass}}
\end{figure}

\begin{figure}
\centering
\includegraphics[width=0.9\textwidth]{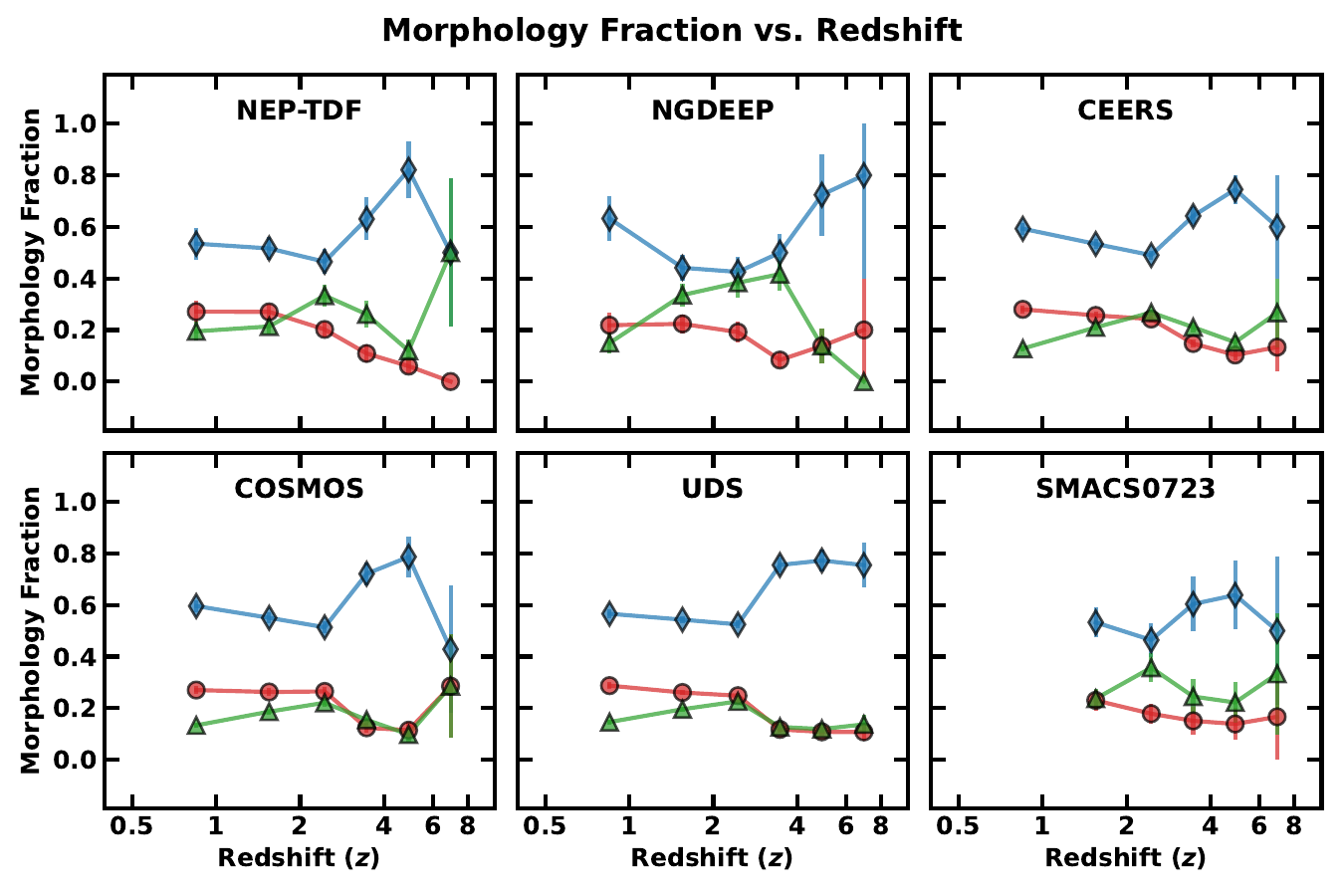}
\caption{
Same as the right panel of {\color{blue} {\bf Figure \ref{fig:mfrac_Mztot}}}, but for each JWST field.
\label{fig:apdx_mfrac_z}}
\end{figure}

The use of six JWST fields with different depths can introduce variations in morphological distributions. 
As shown in {\color{blue} \textbf{Table \ref{tab:zpzs}}}, the outlier fraction of our photometric redshift measurements exhibits fluctuations across the JWST fields, indicating potential variances in the accuracy of stellar masses derived from the SED fitting.
In addition, variations in background fluctuations can lead to systematic differences in the estimation of key parameters ($n$, $A$, and $RFF$) for morphology classification.
To account for these variations, we measured the field-to-field deviations of morphological fractions in the mass and redshift bins, by calculating standard deviations of the fractions in each bin weighted by the number of samples in each field.
The total uncertainties in {\color{blue} \textbf{Figures \ref{fig:mfrac_Mz}, \ref{fig:mfrac_Mztot}, and \ref{fig:mfrac_Mz3}}} were determined by summing the squared field-to-field deviations and Poisson noises of the sample sizes in the bins. 
Here we illustrate the morphological fractions in the six JWST fields as functions of stellar mass and redshift in {\color{blue} \textbf{Figure \ref{fig:apdx_mfrac_mass}}} and {\color{blue} \textbf{Figure \ref{fig:apdx_mfrac_z}}}, respectively.

{\color{blue} \textbf{Figure \ref{fig:apdx_mfrac_mass}}} shows that disk galaxies have a dominant proportion of $\sim60\%$ in ${\rm log}~M_{\ast}/M_{\odot}<10.5$ for all JWST fields.
Although disk fractions generally decrease with increasing stellar mass, this trend is less discernible in the fields of NEP-TDF and NGDEEP due to their limited sample sizes in the mass bin of ${\rm log}~M_{\ast}/M_{\odot}=11.4-12.0$.
Spheroid fractions increase with stellar mass, but the data in the NGDEEP field suffer from the small number statistics in the highest mass bin.
Irregular fractions remain constant at $\sim20\%$, but show large fluctuations in the highest mass bin due to significant uncertainties.
Irregular fractions in the NGDEEP field are higher than in other fields, due to relatively larger $RFF$ values resulting from small background sigma values in Equation \ref{eqn:rff} (see Section \ref{sec:A_RFF}).
Although the data from the SMACS0723 field could be affected by gravitational lensing, 
the general tendency in this field is consistent with the stacked data in the left panel of {\color{blue} \textbf{Figure \ref{fig:mfrac_Mztot}}}.

In {\color{blue} \textbf{Figure \ref{fig:apdx_mfrac_z}}}, disk galaxies also maintain dominance across all redshift ranges in all JWST fields.
Decreases in disk fraction from $\sim70\%$ at $z\gtrsim3$ to $\sim50\%$ at $z\lesssim3$ are consistently observed in all fields.
Despite fluctuations in the disk fraction at the highest redshift bin ($z=6.0-8.0$), the general trend appears unaffected by field-to-field variations.
Spheroid fractions consistently rise with decreasing redshift, except for the cases with only five spheroids in the high-redshift bins ($z=4.0-8.0$) in the NGDEEP field.
Irregular fractions also exhibit similar patterns across all JWST fields, peaking at over $\sim25\%$ around $z\sim2-3$.
Like other morphological types, irregular fractions in the $z=6.0-8.0$ bin are less reliable in the NEP-TDF and NGDEEP fields due to their small sample sizes (only 3 irregulars in the NEP-TDF and none in the NGDEEP).
The NGDEEP field shows notably higher irregular fractions of about $\sim40\%$ at $z=2-3$ compared to other fields, for the reasons mentioned above.

These findings demonstrate that the general trends for each morphological type are not significantly affected by the field-to-field variations.
However, these variations are still non-negligible, so we reflect these fluctuations across the JWST fields as the uncertainties of morphological fractions in the mass and redshift bins.

\section{Conversion from Three-dimensional to Two-dimensional Asymmetry}

\restartappendixnumbering

\begin{figure}
\centering
\includegraphics[width=0.55\textwidth]{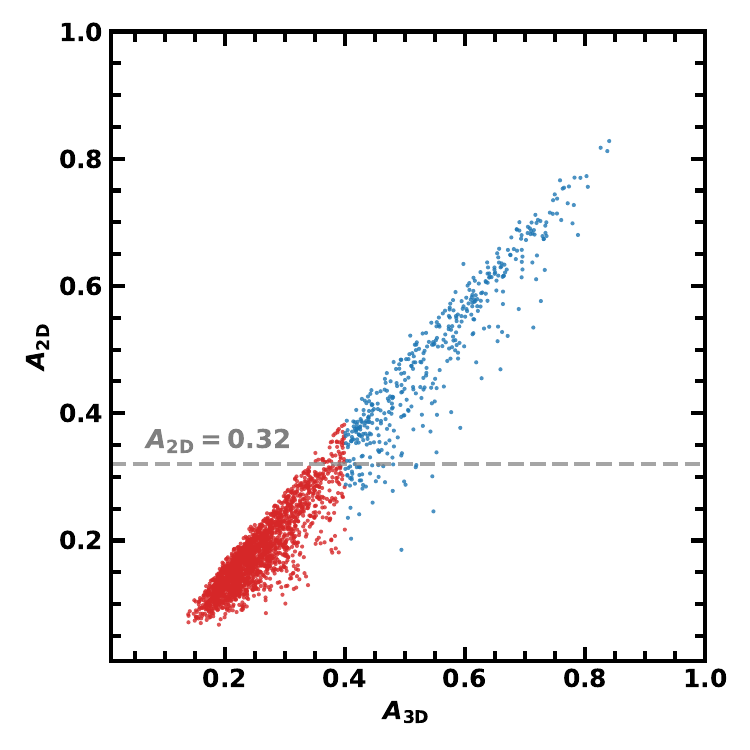}
\caption{
Comparison between the three-dimensional asymmetry ($A_{\rm 3D}$; x-axis) and the two-dimensional asymmetry ($A_{\rm 2D}$; y-axis).
Blue circles denote the objects with $A_{\rm 3D}>0.4$, which are defined as asymmetric galaxies in \citet{park22}, while red circles represent the rests. Note that the division threshold of $A_{\rm 3D}=0.4$ has been found based on the actual distribution $A$ of the galaxies identified in HR5.
Gray dashed line marks the asymmetry criterion of $A_{\rm 2D}=0.32$ used in this study.
\label{fig:apdx_Asym}}
\end{figure}

The asymmetry criterion of $A=0.4$ applied in \citet{park22} to distinguish symmetric galaxies and asymmetric galaxies was established based on the three-dimensional distribution of stellar particles.
Thus, the direct application of this criterion to the two-dimensional data of JWST images is not desirable.
To address this, we conducted a simple test to determine a two-dimensional asymmetry criterion that corresponds to the three-dimensional asymmetry threshold of $A=0.4$ employed based on the actual distribution $A$ of the galaxies identified in HR5.

In {\color{blue} \textbf{Figure \ref{fig:apdx_Asym}}}, we present a comparison between the three-dimensional asymmetry ($A_{\rm 3D}$) and the two-dimensional asymmetry ($A_{\rm 2D}$) of $2,669$ simulated galaxies at $z\sim6$ in the HR5 simulation.
The two-dimensional asymmetry was measured using the same methodology described in Section 3.1 of \citet{park22}, with the face-on projected stellar mass density profile.
We searched for the optimal threshold value of $A_{\rm 2D}>X$ corresponding to $A_{\rm 3D}>0.4$, which minimizes contamination and loss of completeness simultaneously.

Contamination quantifies the presence of contaminated sources within asymmetric objects selected with the $A_{\rm 2D}>X$ criterion, which was calculated as the number fraction of objects with $A_{\rm 2D}>X$ and $A_{\rm 3D}<0.4$ relative to those with $A_{\rm 2D}>X$ and $A_{\rm 3D}>0.4$.
Completeness represents the fraction of objects with $A_{\rm 2D}>X$ and $A_{\rm} > 0.4$ relative to the total number of those with $A_{\rm 3D}>0.4$, so loss of completeness is the opposite probability of the completeness.
We determined the threshold for two-dimensional asymmetry at $X=0.32$ (denoted by the black dashed line in the figure), yielding minimum values for both contamination and loss of completeness about $\sim10\%$.
This two-dimensional asymmetry criterion of $A_{\rm 2D}=0.32$ was subsequently applied in our morphology classification, as detailed in Section \ref{sec:mclass}.

\section{Uncertainties of Morphological Measurements from the PSF Models}

\restartappendixnumbering

\begin{figure}
\centering
\includegraphics[width=0.85\textwidth]{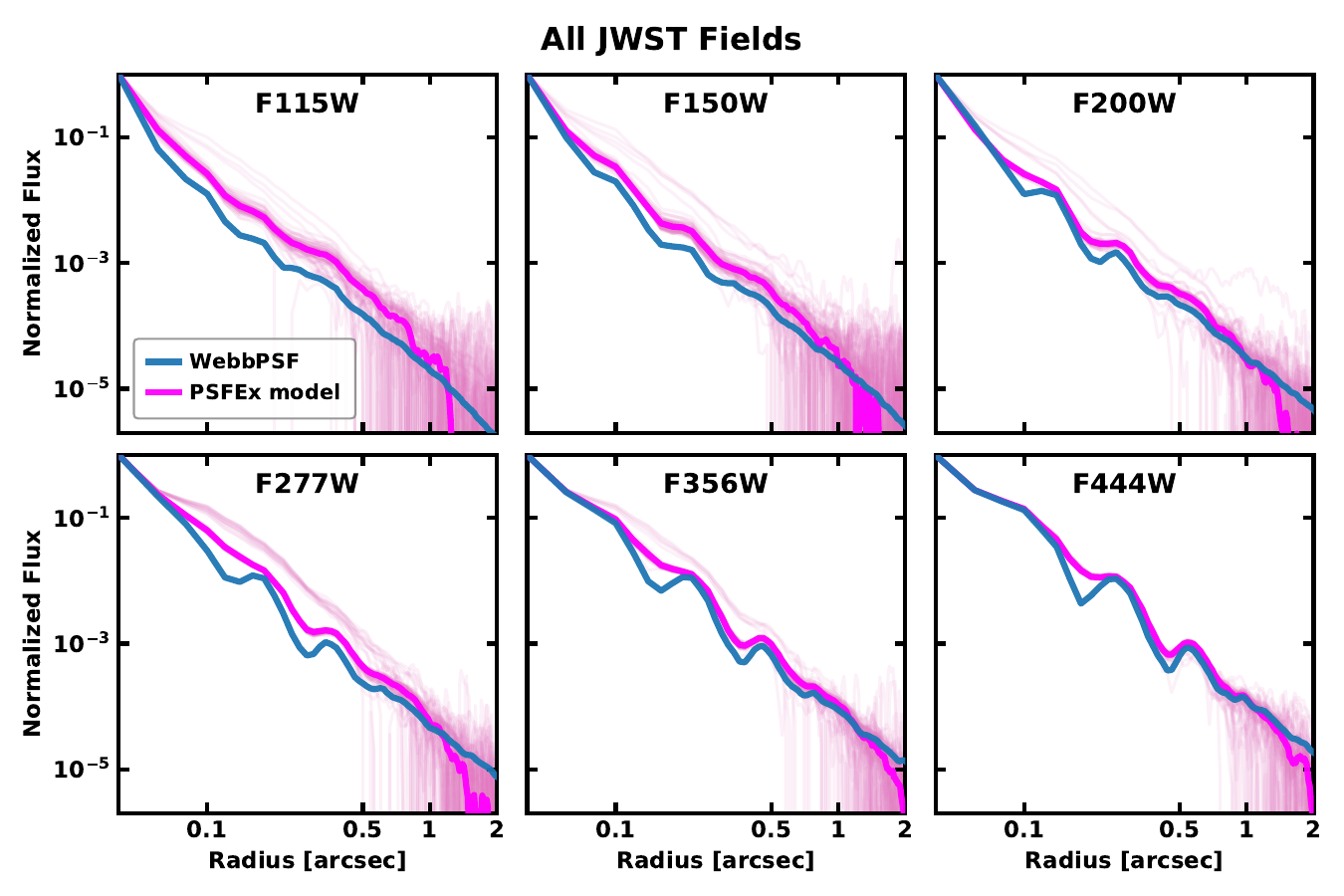}
\caption{
Radial profiles of PSF models for the six JWST/NIRCam bands.
In a panel for each band, the blue thick lines represent the simulated PSF models generated by \textsc{WebbPSF}, and the magenta thick lines represent the median profiles of empirical PSF models from \textsc{PSFEx}.
The pale magenta lines show the profiles of all \textsc{PSFEx} models from each divided region, indicating the spatial variations of empirical PSFs.
\label{fig:apdx_psfprof}}
\end{figure}

\begin{figure}
\centering
\includegraphics[width=0.85\textwidth]{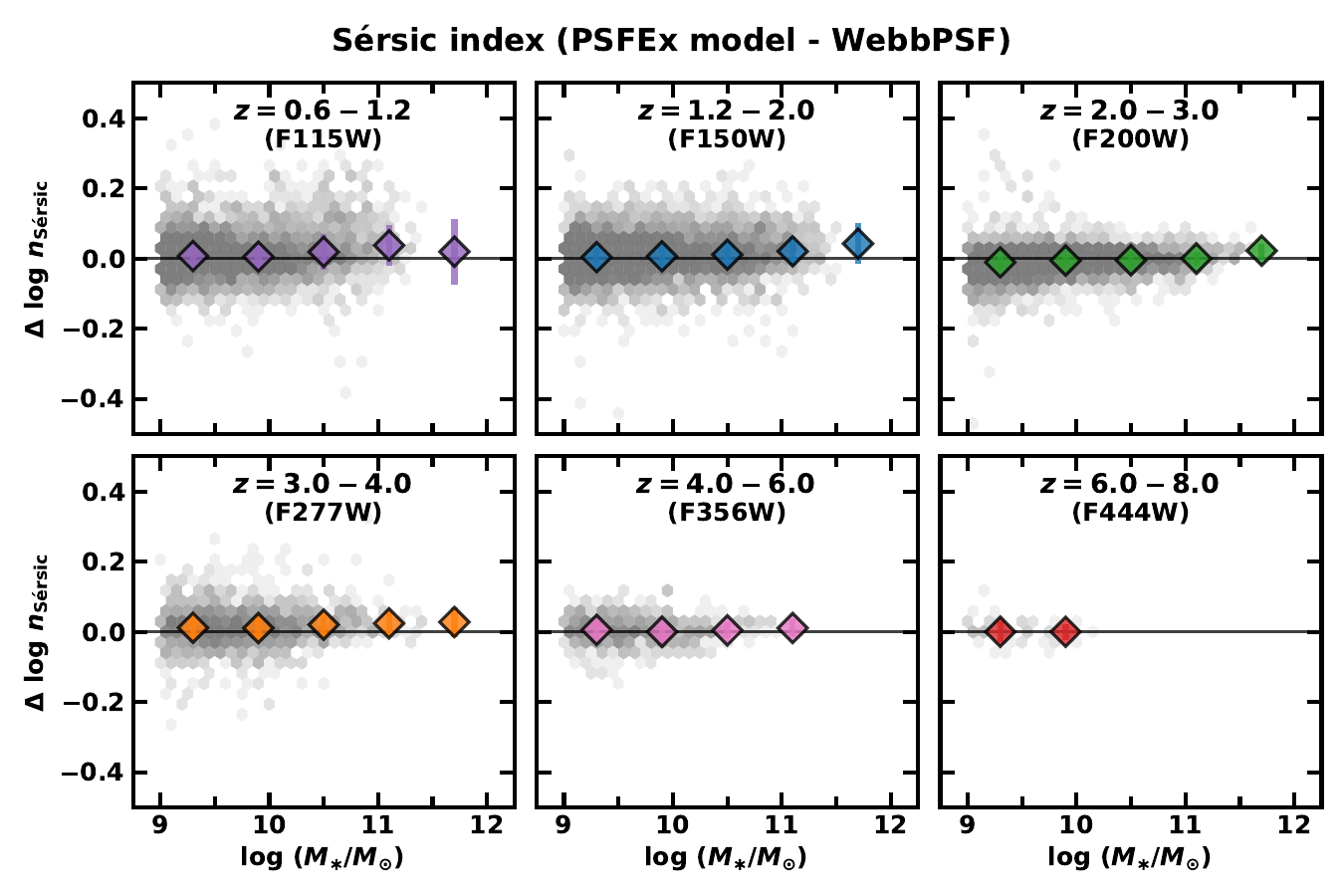}
\caption{
Same as {\color{blue} \textbf{Figure \ref{fig:n_vs_M}}}, but the y-axis value is the logarithmic differences between S\'ersic indices derived from the empirical PSFs (\textsc{PSFEx} model) and those from the simulated PSFs (\textsc{WebbPSF}).
\label{fig:apdx_psfn}}
\end{figure}

\begin{figure}
\centering
\includegraphics[width=0.85\textwidth]{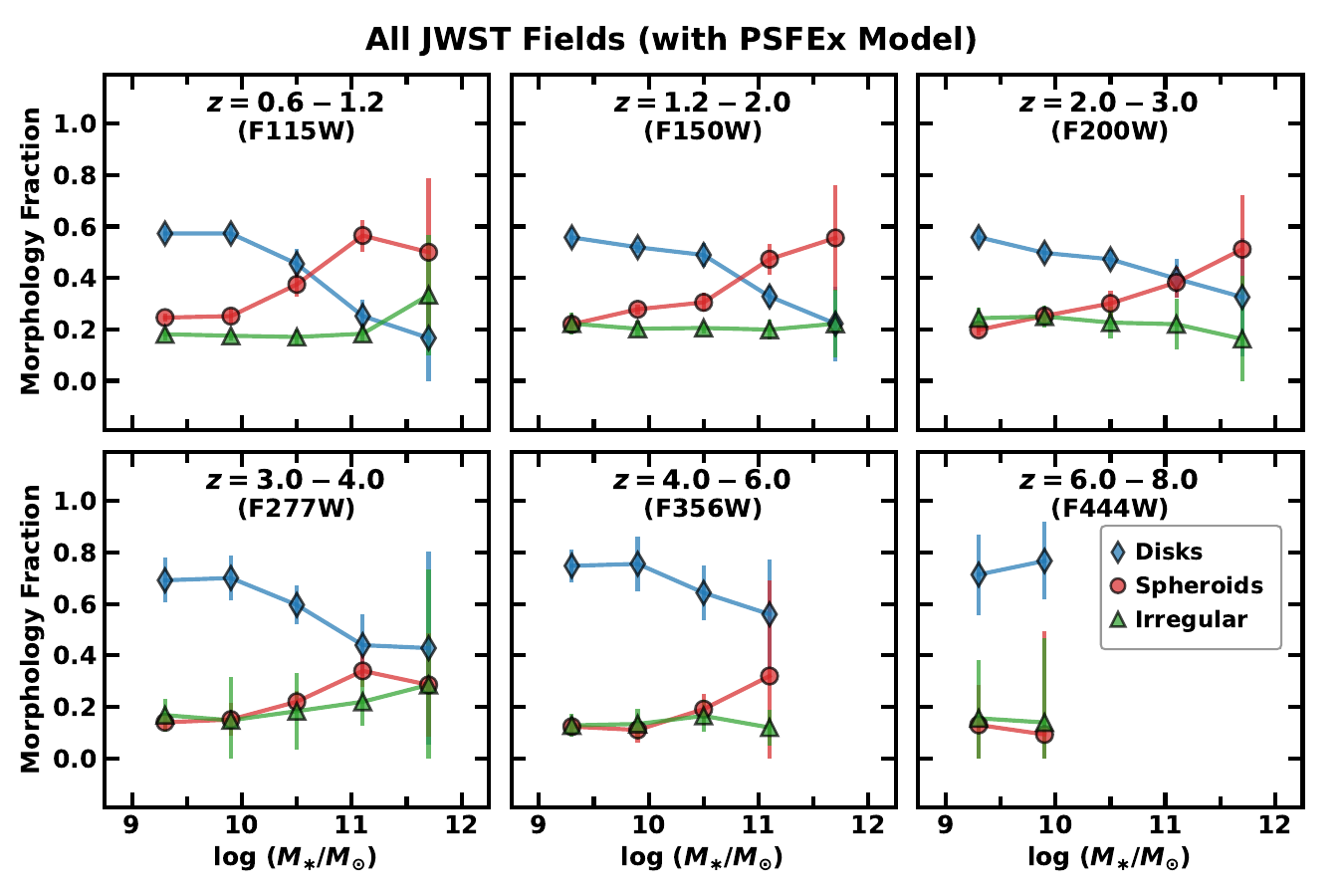}
\caption{
Same as {\color{blue} \textbf{Figure \ref{fig:mfrac_Mz}}}, but the results are derived from the empirical PSF models.
\label{fig:apdx_psf_mfrac}}
\end{figure}

To test the effect of the PSF model for \textsc{GALFIT}, we generated new empirical PSFs with the pixel-based model using the \textsc{PSFEx} software, as done in \citet{zhu24}.
We set our configurations for selecting point sources, with the FWHM range (\texttt{SAMPLE\_FWHMRANGE}) of $1-4$ pixels for short wavelength channels and $2-6$ pixels for long wavelength channels, the minimum signal-to-noise ratio (\texttt{SAMPLE\_MINSN}) of 100, and the maximum $(a-b)/(a+b)$ value (\texttt{SAMPLE\_MAXELLIP}) of 0.15.
The output PSF models are sampled with the pixel scale of $0\farcs04~{\rm pixel}^{-1}$, with the image size of $4\arcsec\times4\arcsec$.
The PSF image size was set to be slightly smaller than that of \textsc{WebbPSF} model ($5\arcsec\times5\arcsec$) due to possible light contamination from neighboring objects in real images.
To account for spatial variations in PSFs, we manually divided the images of JWST fields into multiple regions, each of which encompasses the field of view of $\sim2\farcm2\times2\farcm2$.
From each region, we extracted the empirical PSF models from $\sim5-15$ selected stars.

{\color{blue} \textbf{Figure \ref{fig:apdx_psfprof}}} shows the radial light profiles of all PSF models.
In all JWST/NIRCam bands, the empirical PSFs from \textsc{PSFEx} have broader profiles compared to the simulated PSF from \textsc{WebbPSF}.
This is also consistent with the profiles of empirical PSFs used in other JWST studies \citep{ono23} because the drizzling effect of real NIRCam images can be reflected in empirical PSFs in contrast to simulated PSFs.
{\color{blue} \textbf{Figure \ref{fig:apdx_psfn}}} shows the differences of S\'ersic indices of our sample derived from \textsc{PSFEx} models and those from \textsc{WebbPSF}.
Across all NIRCam bands, the S\'ersic index differences show considerable scatters about $\sim0.1~{\rm dex}$ but little systematic biases.
The S\'ersic indices from the \textsc{PSFEx} models can be slightly increased in the high-mass regime compared to those from \textsc{WebbPSF} because deconvolving galaxies with broader PSFs in \textsc{GALFIT} tend to result in more concentrated galaxy models with larger S\'ersic indices.
However, the median increases of the S\'ersic indices are less than $\sim0.04$ dex, leading to little systematic biases.

Applying the \textsc{PSFEx} models, {\color{blue} \textbf{Figure \ref{fig:apdx_psf_mfrac}}} shows the results of morphological fractions as a function of stellar mass.
Comparing this figure to {\color{blue} \textbf{Figure \ref{fig:mfrac_Mz}}}, the overall trends of morphological fractions do not change at all, still supporting our results.
Thus, the choice of the PSF models does not have significant systematic effects on our main conclusions.


    
\end{document}